\documentclass[%
 reprint,
superscriptaddress,
 amsmath,amssymb,
 aps,
]{revtex4-2}

\usepackage{graphicx}%
\usepackage{dcolumn}%
\usepackage{bm}%
\usepackage{hyperref}%
\hypersetup{
  final,
  pdfborder={0 0 0},
  colorlinks=true,
  urlcolor=blue,
  linkcolor=blue,
  citecolor=blue
}
\usepackage{siunitx}
\DeclareSIUnit\angstrom{\text{Å}} %
\usepackage{miller}
\usepackage{mathtools}
\usepackage{esvect}
\usepackage{float}
\begin{document}
\frenchspacing

\title{Effect of the atomic structure of complexions on the active disconnection mode during shear-coupled grain boundary motion} %

\author{Swetha Pemma}
\affiliation{%
     Max-Planck-Institut für Eisenforschung GmbH, Max-Planck-Stra\ss{}e 1, 40237 D\"usseldorf, Germany
}%
\author{Rebecca Janisch}%
\affiliation{%
 Interdisciplinary Centre of Advanced Materials Simulation (ICAMS), Ruhr-Universität Bochum, 44780 Bochum, Germany\\
}%
\author{Gerhard Dehm}%
\email{dehm@mpie.de}
\affiliation{%
     Max-Planck-Institut für Eisenforschung GmbH, Max-Planck-Stra\ss{}e 1, 40237 D\"usseldorf, Germany
}%
\author{Tobias Brink}%
\email{t.brink@mpie.de}
\affiliation{%
     Max-Planck-Institut für Eisenforschung GmbH, Max-Planck-Stra\ss{}e 1, 40237 D\"usseldorf, Germany
}%

\date{\today}%

\begin{abstract}
The migration of grain boundaries leads to grain growth in polycrystals and is one mechanism of grain-boundary-mediated plasticity, especially in  nanocrystalline metals. This migration is due to the movement of dislocation-like defects, called disconnections, which couple to externally applied shear stresses. While this has been studied in detail in recent years, the active disconnection mode was typically associated with specific macroscopic grain boundary parameters. We know, however, that varying microscopic degrees of freedom can lead to different atomic structures without changing the macroscopic parameters. These structures can transition into each other and are called complexions. Here, we investigate $[11\overline{1}]$ symmetric tilt boundaries in fcc metals, where two complexions---dubbed domino and pearl---were observed before. We compare these two complexions for two different misorientations: In $\Sigma19$b $[11\overline{1}]$ $(178)$ boundaries, both complexions exhibit the same disconnection mode. The critical stress for nucleation and propagation of disconnections is nevertheless different for domino and pearl. At low temperatures, the Peierls-like barrier for disconnection propagation dominates, while at higher temperatures the nucleation is the limiting factor. For $\Sigma$7 $[11\overline{1}]$ $(145)$ boundaries, we observed a larger difference. The domino and pearl complexions migrate in different directions under the same boundary conditions. While both migration directions are possible crystallographically, an analysis of the complexions' structural motifs and the disconnection core structures reveals that the choice of disconnection mode and therefore migration direction is directly due to the atomic structure of the grain boundary. 
\end{abstract}

\maketitle

\newcounter{supplfigctr}
\renewcommand{\thesupplfigctr}{S\arabic{supplfigctr}}
\newcounter{suppltabctr}
\renewcommand{\thesuppltabctr}{S\arabic{suppltabctr}}
{\refstepcounter{supplfigctr}\label{fig:convergefit}}
{\refstepcounter{suppltabctr}\label{tab:energies_parameters}}
{\refstepcounter{suppltabctr}\label{tab:Ecore}}
{\refstepcounter{supplfigctr}\label{fig:S19b_burgers_circuit}}
{\refstepcounter{supplfigctr}\label{fig:S7_burgers_circuit}}
{\refstepcounter{supplfigctr}\label{fig:Emax}}
{\refstepcounter{supplfigctr}\label{fig:shearstressT}}
{\refstepcounter{supplfigctr}\label{fig:velocity_beta_temp}}

\section{\label{sec:intro}Introduction}

Grain boundaries (GBs) influence mechanical properties of polycrystalline materials and GB engineering is critical in materials design \cite{Randle_2010}. The motion of GBs is the key factor in the microstructure evolution of poycrystalline materials \cite{Simpson_1971, Riontino_1979}. When subjected to shear stress, GBs move and can account for part of the plastic deformation in nanocrystalline materials \cite{Van_Swygenhoven_2001, Shan_2004, Meyers2006, Lohmiller2014}.
Some of the GB-related deformation mechanisms discussed in the literature are GB sliding \cite{FUKUTOMI1985, FUKUTOMI1991, Langdon_2006, Wang_2022}, grain rotation \cite{Murayama_2002,Cahn2004, Upmanyu_2006, Wang_2014, Thomas2017}, shear-coupled GB migration \cite{Winning2001, Winning2002, Suzuki_2005, Cahn2006, Gianola2006, Mishin2007, Ivanov2008, Gorkaya_2009, Caillard2009, Mompiou2009, Mompiou2011,Molodov2011, Rajabzadeh2013, Li2013, Homer2013, Rajabzadeh2014, Molodov2018,Zhu_2019, Wei2021}, diffusional creep \cite{Meyers2006, Dao2007}, dislocation interaction at GBs \cite{Van_Swygenhoven_2001, Yamakov_2002, Meyers2006, Dao2007}, and enhanced partial dislocation activity \cite{Van_Swygenhoven_2001, Yamakov_2002, Meyers2006, Dao2007}.

Shear coupling is the migration of GBs driven by shear stress across the GB plane \cite{Hirth1996, Cahn2006, Caillard2009}. It can lead to complex effects during grain growth in a polycrystal, such as grain rotation, stress generation, and grain growth stagnation, which are all inter-related \cite{Thomas2017}. A shear coupling factor $\beta = v_\parallel / v_\perp$ describes how a GB migrates: A relative shear velocity $v_\parallel$ of the two grains parallel to the GB is coupled directly to the GB migration velocity $v_\perp$ normal to its plane \cite{Ashby_1972, Cahn2004, Cahn2006, Han2018}. This factor $\beta$ is influenced by parameters such as temperature, bi\-crys\-tal\-logra\-phy, and the type of the driving force \cite{Chen_2019}. The existing models of conservative GB kinetics, such as the GB dislocation model \cite{Cahn2004, Cahn2006}, the shear migration geometrical model \cite{Mompiou2010,Caillard2009}, and GB disconnections \cite{King1980,Rae1980, Hirth1996, Ashby_1972,Cahn2006,Han2018} are used to explain the effect of GB geometry on shear-coupled motion. Microscopically, shear-coupled motion is caused by the movement of disconnections, which are line defects at the GB \cite{Bollmann_1970, Ashby_1972, Hirth_1973, Han2018}. Disconnections have dislocation character insofar that they have a Burgers vector $\mathbf{b}$, which couples to externally applied stress. They also lead to a step of size $h$ in the GB, which results in the GB migration normal to its habit plane during disconnection nucleation and movement \cite{Hirth1996, Hirth_2006}. The shear coupling factor $\beta$ arises directly from the coupling of \textbf{b} to $h$. Consequently, the formation and migration of disconnections play a vital role in the kinetic properties of GBs \cite{Rajabzadeh2013a, Combe2016, MacKain2017, Chen_2020, Combe2021} and pre-existing mobile disconnections lead to a reduced stress required for GB migration \cite{Khater2012}. 

There are several possible disconnection modes (combinations of \textbf{b} and $h$) that can be activated during shear-coupled motion. Prior works have considered the choice of active mode based on macroscopic GB parameters (bicrystallography), as well as magnitude of applied stress and temperature \cite{Cahn2006, Combe2016, Molodov2018, Han2018, Chen_2019, Chen_2020, Chen_kt_2020, Chesser_2021}. However, even GBs with the same macroscopic parameters can exhibit different atomic structures due to the microscopic degrees of freedom. Such differences in atomic GB structure and first-order transitions between them were observed even in pure metallic materials \cite{Frolov2013, Aramfard2018, Zhu2018, Frolov2018, Yang2020, Meiners2020, Langenohl2022, Brink_2023}. These different structures can be treated as interface phases, which can only exist in contact with the abutting crystallites, and can be treated using a thermodynamic framework \cite{GibbsVol1, Hart1968, Cahn1982, Rottman1988, Frolov2012, Frolov2012a}. They are called complexions \cite{Tang2006, Dillon2007, Cantwell2014, Cantwell2020} or GB phases \cite{Frolov2015}. Complexion transitions, then, are analogous to bulk phase transitions: The GB structure, composition, and properties change discontinuously at critical values of thermodynamic parameters such as temperature, pressure, and chemical potential \cite{Tang2006, Frolov2012, Frolov2012a, Frolov2015, Cantwell2014, Cantwell2020}. Until now, there is only one study \cite{Frolov2014} that reports that two complexions in $\Sigma5$ [001] (210) tilt GBs have different signs of the shear coupling factor. This emphasizes that complexions can affect the choice of active shear-coupling mode, but it was not reported why this occurs. Furthermore, that study only covers a single GB and more evidence would be needed to reveal if this is a common phenomenon or an exceptional case.

In this paper, we thus report on the shear-coupled motion of $\Sigma19$b $[11\overline{1}]$ $(178)$ and $\Sigma7$ $[11\overline{1}]$ $(145)$  symmetric tilt GBs, which are of interest because they both have two possible complexions, named domino and pearl \cite{Meiners2020, Langenohl2022, Brink_2023}. We find that the complexions of the $\Sigma7$ GB also exhibit different shear-coupling factors $\beta$ and explain how the choice of $\beta$ is due to their structural motifs. The $\Sigma19$b complexions both have the same $\beta$, but vary in the critical shear stress needed to activate the shear-coupled motion, which we investigate in detail.

\section{\label{sec:method}Computational methods}

We studied bicrystals with symmetric tilt GBs using molecular dynamics (MD) simulations, which were performed using LAMMPS \cite{Plimpton1995, Thompson_2022} with the embedded atom method (EAM) potential of Mishin et al. \cite{Mishin2001}. This potential reproduces some properties of Cu very well, comprising elastic constants, phonon frequencies, thermal expansion, intrinsic stacking fault energy, the coherent twin boundary energy, and others. We used a time integration step of \SI{2}{fs} for all dynamics simulations.

\begin{figure}[t!]
\includegraphics{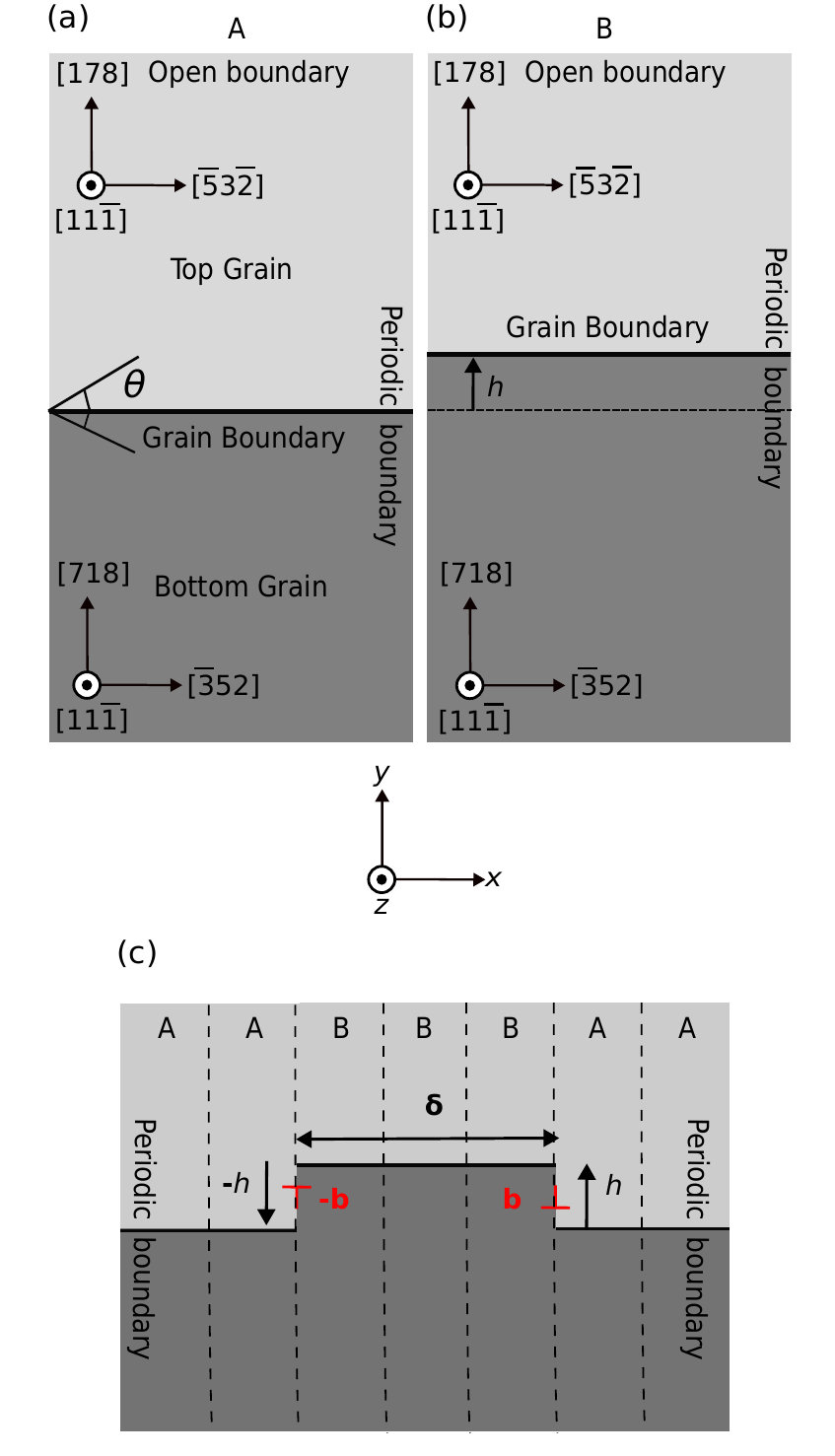}
\caption{\label{fig:GB_sketch}(a) Schematic of the bicrystallography of a $\Sigma$19b $[11\overline{1}]$ $(178)$ symmetric tilt GB, marked as bicrystal A. Top and bottom fcc crystals are joined in the indicated orientations, leading to a misorientation between the two crystals of \ang{46.83}. Similar GB construction followed for $\Sigma$7 $[11\overline{1}]$ $(145)$ symmetric tilt GB (Table~\ref{tab:GBparameters}). (b) Bicrystal B has a GB that is moved by one migration step compared to bicrystal A. This was achieved by moving the GB plane by a step height and shifting the top crystal by the corresponding Burgers vector before minimization. (c) Schematic of the construction of the disconnection dipole with opposite Burgers vectors by assembling a simulation cell from bicrystals A and B.  By varying the number of B unit cells, the dipole width $\delta$ between disconnections can be controlled. Periodic boundary conditions were applied in  $x$ and $z$ directions and open boundary conditions in $y$ direction.
}
\end{figure}

The bicrystals for $\Sigma$19b $[11\overline{1}]$ $(178)$ symmetric tilt GBs (misorientation of \ang{46.83}) were created by constructing two fcc crystals with the desired crystallographic orientations $[\overline{5}3\overline{2}]$, $[178]$, $[11\overline{1}]$ in $x$, $y$, $z$ directions for the top crystal and $[\overline{3}52]$, $[718]$, $[11\overline{1}]$ for the bottom crystal [Fig.~\ref{fig:GB_sketch}(a)]. Here, $y$ is the GB normal and $z$ the tilt axis. The $x$ and $z$ directions were periodic and we used open boundaries in $y$ direction, allowing us to produce differently-sized cells by repeating the unit cell along the periodic directions. Similarly, the bicrystals for $\Sigma$7 $[11\overline{1}]$ $(145)$ symmetric tilt GBs (misorientation of \ang{38.21}) were created by constructing two fcc crystals with the desired crystallographic orientations $[\overline{3}2\overline{1}]$, $[145]$, $[11\overline{1}]$ in $x$, $y$, $z$ directions for the top crystal and $[\overline{2}31]$, $[415]$, $[11\overline{1}]$ for the bottom crystal (Table~\ref{tab:GBparameters}).

The GB structures were formed by combining the two crystallites, sampling the microscopic degrees of freedom by displacing the top crystal, and minimizing the result in molecular statics ($\gamma$-surface method). We did this until the previously reported pearl and domino complexions  for both the GBs \cite{Meiners2020, Langenohl2022, Brink_2023} were found (we evaluated that structure, GB energy, and excess volume match). The minimum energy GB structures for domino and pearl are visualized using OVITO \cite{Stukowski2009} and are shown in Fig.~\ref{fig:shearcoupling}.

The periodic unit cell dimensions of the $\Sigma$19b GB were $L_x = \SI{11.142}{\angstrom}$ and $L_z = \SI{6.261}{\angstrom}$. We used $L_y$ ranging from $\SI{385.317}{\angstrom}$ to $\SI{5790.070}{\angstrom}$ (see below). The periodic unit cell dimensions of the $\Sigma$7 GB were $L_x = \SI{6.763}{\angstrom}$ and $L_z = \SI{6.261}{\angstrom}$, with $L_y$ ranging from $\SI{210.487}{\angstrom}$ to $\SI{1171.61}{\angstrom}$. These unit cells were used as building blocks that can be repeated along the periodic $x$ and $z$ directions.

\begin{table}
\caption{\label{tab:GBparameters} Symmetric $\Sigma$ tilt GBs used to study shear coupled GB motion.}
\begin{ruledtabular}
\begin{tabular}{lcccc}
CSL type &  $\Sigma$7 & $\Sigma$19b\\
 \hline
Misorientation & \ang{38.21} & \ang{46.83} \\ 
Tilt axis  & $[11\overline{1}]$ & $[11\overline{1}]$\\
GB planes & $(145)$ $(415)$ &  $(178)$ $(718)$\\
\end{tabular}
\end{ruledtabular}
\end{table}

For the simulation of shear-coupled GB motion of $\Sigma$19b, we used a simulation cell of size $222.814 \times 385.317
\times \SI{62.613}{\angstrom^3}$ ($20 \times 20 \times 10$ unit cells, resulting in 455,400 atoms) unless specified otherwise. For $\Sigma$7, a simulation cell of size of $135.258 \times 210.487
\times \SI{62.613}{\angstrom^3}$ ($20 \times 18 \times 10$ unit cells, resulting in 151,200 atoms) was used.
We started by using molecular statics simulations ($T = \SI{0}{K}$) and applying a displacement on the top boundary in $x$ direction and keeping the bottom boundary fixed. Both boundaries were the regions at the surfaces (in $y$ direction). Each boundary region is with an extent of \SI{55}{\angstrom} for the $\Sigma$19b GBs and an extent of \SI{30}{\angstrom} for the $\Sigma$7 GBs (Fig.~\ref{fig:shearcoupling}). 
To study GB migration in domino and pearl, the shear displacement $d$ was applied stepwise in increments of \SI{0.05}{\angstrom}, while minimizing the system after every step.
We calculated the resulting shear stress by dividing the reaction force at the boundary by the area $L_x \times L_z$ of the top boundary.

Then, MD simulations were performed in the canonical ensemble (Nos\'e--Hoover thermostat at $T = 100$, 200, 300, 400, 500, \SI{600}{K}). At finite temperatures, the system was scaled to obtain the correct lattice constant at the desired temperature $T$ and then equilibrated for \SI{4}{ns}. In this simulation procedure, we applied a velocity in $x$ direction to the top boundary, while keeping the bottom boundary fixed (see Ref.~\cite{Cahn2006} and Fig.~\ref{fig:shearcoupling}). The top boundary was allowed to move freely in $y$ and $z$ direction. Shear-coupled simulations were performed for shear velocities in the range from \SI{0.01}{m/s} to \SI{10}{m/s}. We typically applied a constant shear velocity of \SI{0.1}{m/s} unless otherwise noted. Shear stress was calculated as for the molecular statics simulations.

Disconnection dipoles observed in both the complexions during shear-coupled motion were constructed manually to obtain disconnection formation and migration energies. The disconnection dipoles were constructed for different dipole widths $\delta$ (separation distance between the two disconnections). This is similar to the construction of disconnections as described in Refs.~\cite{Winter2022, Winter_phase_2022}: Bicrystals A and B were generated such that they contain the same complexion [Fig.~\ref{fig:GB_sketch}(a)--(b)]. The GB in bicrystal B was moved by one step height compared to A. By replicating A and B and assembling them (e.g., as AABBBAA), two disconnections with opposite Burgers vectors appear at the junctions \ldots{}AB\ldots{} and \ldots{}BA\ldots{} [Fig.~\ref{fig:GB_sketch}(c)]. By varying the number of B unit cells, the separation $\delta$ between disconnections can be controlled. We calculated the energy of the disconnection dipoles in simulation cells of sizes $1114.220 \times L_y \times \SI{6.261}{\angstrom^3}$ (100 unit cells in $x$ direction). To ensure convergence with $L_y$, we used values from $L_y = \SI{964.490}{\angstrom}$ (\SI{5.6e5}{atoms}) to \SI{5790.070}{\angstrom} (\SI{3.4e6}{atoms}). We found that the dipole energies converged at $L_y = \SI{1929.540}{\angstrom}$, corresponding to $100\times100\times1$ unit cells with {\SI{1.1e6}{atoms}} (see Fig.~\ref*{fig:convergefit} in the supplemental material (SM) \cite{suppl}). For $\Sigma$7, we also used $100\times100\times1$ unit cells, resulting in simulation cells of size $676.304 \times 1171.61 \times \SI{6.261}{\angstrom^3}$ with \SI{4.2e5}{atoms}.

The energy of the resulting disconnection dipole is
\begin{equation}
    E_\text{dipole} = \frac{E_1 - E_0}{L_z},
\end{equation}
where $E_0$ is the energy of a system with a GB but no disconnection, $E_1$ is the energy of the same system with a disconnection dipole, and $L_z$ is the width of the system along the disconnection lines. The dipole energy consists of core energies for each disconnection and an elastic interaction energy \cite{Nabarro1952, Hirth1983, Khater2012, Han2018}:
\begin{equation}
\label{eq:Edipole}
 E_{\text{dipole}} = 2E_{\text{core}} +  E_{\text{elastic}}.
\end{equation}
Sometimes the core energy is split into a core and step energy, where the energy cost of the GB step is considered separately. Here, we will consider the core and step energy in terms of a single core energy. The elastic interaction energy is \cite{Hirth1983}
\begin{equation}
E_{\text{elastic}} = K b^2 \ln\left(  \frac{  \delta } { \delta_{c} }\right),
\label{eq:E_nonper}
\end{equation}
where $K$ is the energy coefficient describing the anisotropic crystal elasticity  \cite{Hirth1983} and $\delta_{c}$ is the disconnection core size. Eqs.~\ref{eq:Edipole} and \ref{eq:E_nonper} can be simplified by a mathematical trick: we can ``hide'' the core energies by defining $\delta_0 = \delta_c \exp (-2E_\text{core}/(Kb^2))$ and writing
\begin{equation}
E_{\text{dipole}} = K b^2 \ln\left(  \frac{  \delta } { \delta_0 }\right).
\label{eq:E_tot_short}
\end{equation}
The length $\delta_0$ is now an effective core size without direct physical meaning, but Eq.~\ref{eq:E_tot_short} can be fitted directly to the dipole energies obtained by molecular statics simulations without knowledge of $E_\text{core}$. With periodic boundary conditions along the $x$ and $z$ directions in a bicrystal simulation, the energy of pair of disconnections is given by \cite{Rajabzadeh2013, MacKain2017, Han2018}
\begin{equation}
E_{\text{dipole}} =  K b^2 \ln\left(  \frac{ \sin(\pi \delta / L_{x}) }{ \sin(\pi \delta_{0} / L_{x}) }\right),
\label{eq:E_delta}
\end{equation}
taking into account the image interactions. When $L_{x}$ is infinite, Eq.~\ref{eq:E_delta} reduces to Eq.~\ref{eq:E_tot_short}. In our simulations of the $\Sigma$19b GBs, we considered the periodic case with a simulation box size of $\SI{1114.220}{\angstrom} \times L_y \times \SI{6.261}{\angstrom}$. We varied $L_y$ to verify that there are no size effects for $L_y \geq \SI{1929.540}{\angstrom}$ (Fig.~\ref*{fig:convergefit}, Tables~\ref*{tab:energies_parameters} and \ref*{tab:Ecore} in the SM \cite{suppl}).

The energy of the disconnection dipole is maximum at the disconnection dipole width $\delta^* = L_x/2$, and its value is
\begin{equation}
  E_{\text{dipole}}^{*}  \approx   K b^2 \ln\left(  \frac{  L_{x} }{ \pi \delta_{0} }\right)
\label{eq:E_max}
\end{equation}
for periodic boundary conditions and $\delta_0/ L_{x} \ll 1$.

Finally, in order to obtain the barriers for the migration of the disconnections themselves, the minimum energy path for a change of the disconnection dipole width $\delta$ was obtained by nudged elastic band (NEB) method calculations \cite{Henkelman2000, Henkelman_2000}. The spring constants for the parallel and perpendicular nudging forces were both $\SI{1.0}{eV/\angstrom^2}$. The minimization scheme used was quickmin \cite{Sheppard_2008}. The initial and final states of the minimum energy path are the GB structures with disconnection dipole widths $\delta$ varying by $\SI{11.142}{\angstrom}$ and $\SI{6.763}{\angstrom}$ for $\Sigma$19b and $\Sigma$7, respectively (equivalent to a CSL periodicity vector and thus the distance between two local minima for the disconnections). The saddle point observed along the minimum energy path is the required disconnection migration barrier (difference in minimum and maximum energy observed along minimum energy path). 

Raw data for all simulations and analyses is available in the companion dataset \cite{zenodo}.

\section{\label{sec:results}Results and Discussion}

\subsection{\label{sec:shear_coupling}Shear coupling factor}

\begin{figure*}
\includegraphics{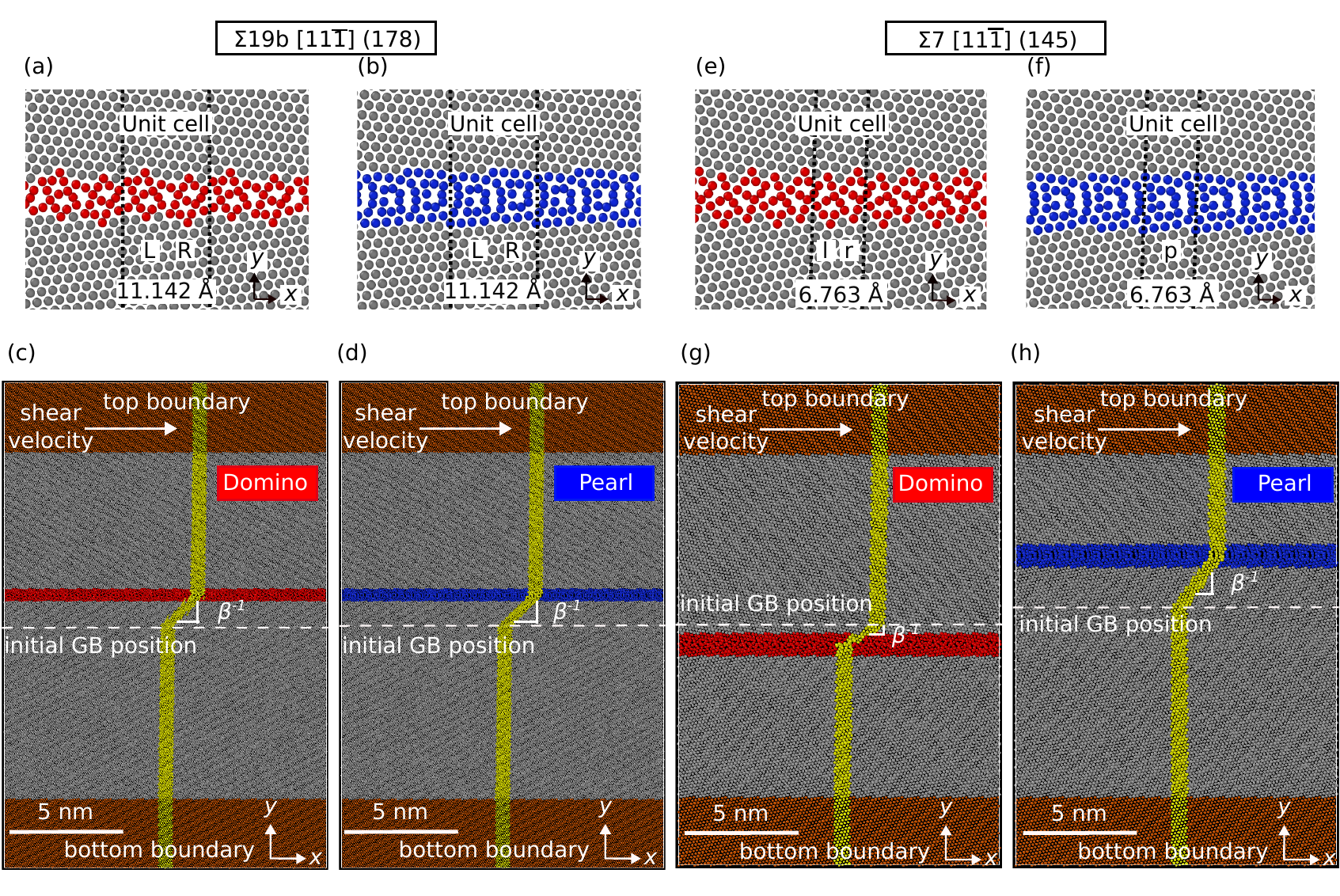}
\caption{\label{fig:shearcoupling} Atomic structures of the (a) domino and (b) pearl complexions of $\Sigma$19b $[11\overline{1}]$ $(178)$ symmetric tilt GBs in our computer model. Both complexions consist of two repeating structural units (\ldots{}LRLRLR\ldots{}). Shear coupling simulations for \SI{20}{ns} with a shear velocity of \SI{0.1}{m/s} at \SI{300}{K} lead to a migration distance on the order of \SI{2}{nm} in the positive $y$ direction for both (c) domino and (d) pearl. Atomic structures of the (e) domino and (f) pearl complexions of $\Sigma$7 $[11\overline{1}]$ $(145)$ symmetric tilt GB in our computer model. Domino consist of two repeating overlapped structural units (\ldots{}lrlrlr\ldots{}). Pearl instead consist of a single repeating structural unit (\ldots{}ppp\ldots{}). Shear coupling simulations for \SI{16}{ns} lead to a migration distance on the order of \SI{0.8}{nm} in the negative $y$ direction  and \SI{2.1}{nm} in the positive $y$ direction for (g) domino and (h) pearl, respectively. The red and blue atoms belong to the GBs, the gray atoms are fcc atoms, and the brown atoms represent the boundary. Shear-coupled GB motion is highlighted by the yellow fiducial mark, which exhibits a slope (equivalent to $\beta^{-1}$) in the region traversed by the GB. The initial GB positions are indicated by the dashed lines.}
\end{figure*}

We investigated $\Sigma$19b $[11\overline{1}]$ $(178)$ and $\Sigma$7 $[11\overline{1}]$ $(145)$ symmetric tilt GBs in copper, which each exhibit both a pearl and a domino complexion (Fig.~\ref{fig:shearcoupling}). These two complexions in the $\Sigma19$b GB can each be thought of as consisting of two alternating motifs, designated L and R [Fig.~\ref{fig:shearcoupling}(a)--(b)]. 
We simulated their shear-coupled motion by applying a shear displacement $d$. The result for the applied shear velocity $v_\parallel = \SI{0.1}{m/s}$ after a simulation time of \SI{20} {ns} at \SI{300}{K} is shown in Fig.~\ref{fig:shearcoupling}(c)--(d). Before the simulations, atoms in a vertical line were marked in yellow. This fiducial line highlights the atomic displacements: The GB has moved in positive $y$ direction [Fig.~\ref{fig:shearcoupling}(c)--(d)] from its original position (dashed line), while the material was sheared in positive $x$ direction. The macroscopically applied displacement thus couples to the GB migration. The slope of the fiducial line provides the ratio of GB migration distance to sliding and is therefore equivalent to the inverse of the shear coupling factor \cite{Ashby_1972,Cahn2004, Cahn2006, Han2018}
\begin{equation}
  \beta_{\Sigma19\mathrm{b}} = \frac{v_\parallel}{v_\perp} = 0.865,
\label{eq:beta}
\end{equation}
where $v_\parallel$ is the shear velocity applied to the system and $v_\perp$ the velocity of GB migration. A more exact calculation method for $\beta$ is provided in Appendix~\ref{sec:shear_coupling_complexions}. We note that both complexions exhibit the same value of $\beta_{\Sigma19\mathrm{b}}$.

For the $\Sigma$7 GBs, we also find domino and pearl structures [Fig.~\ref{fig:shearcoupling}(e)--(f)], but with some differences to the $\Sigma19$b GBs. The domino complexion also consists of two alternating motifs, but they partially overlap. We therefore designate these motifs as lower-case l and r. The pearl complexion, however, only has a single motif per unit cell (here designated p). The results of the shear-coupling simulations are shown in Fig.~\ref{fig:shearcoupling}(g)--(h). We note that the two complexions migrate in opposite directions under the same applied boundary conditions and obtain (see also Appendix~\ref{sec:shear_coupling_complexions})
\begin{align}
    \beta_{\Sigma7}^\text{pearl} &= +0.692 \quad \text{and}\\
    \beta_{\Sigma7}^\text{domino} &= -1.734.
\end{align}
To understand this, we need to look more closely at the GB migration mechanism.

\subsection{\label{sec:bicrystallography}Origin of the active disconnection modes}

\begin{figure*}
\includegraphics[scale=0.92]{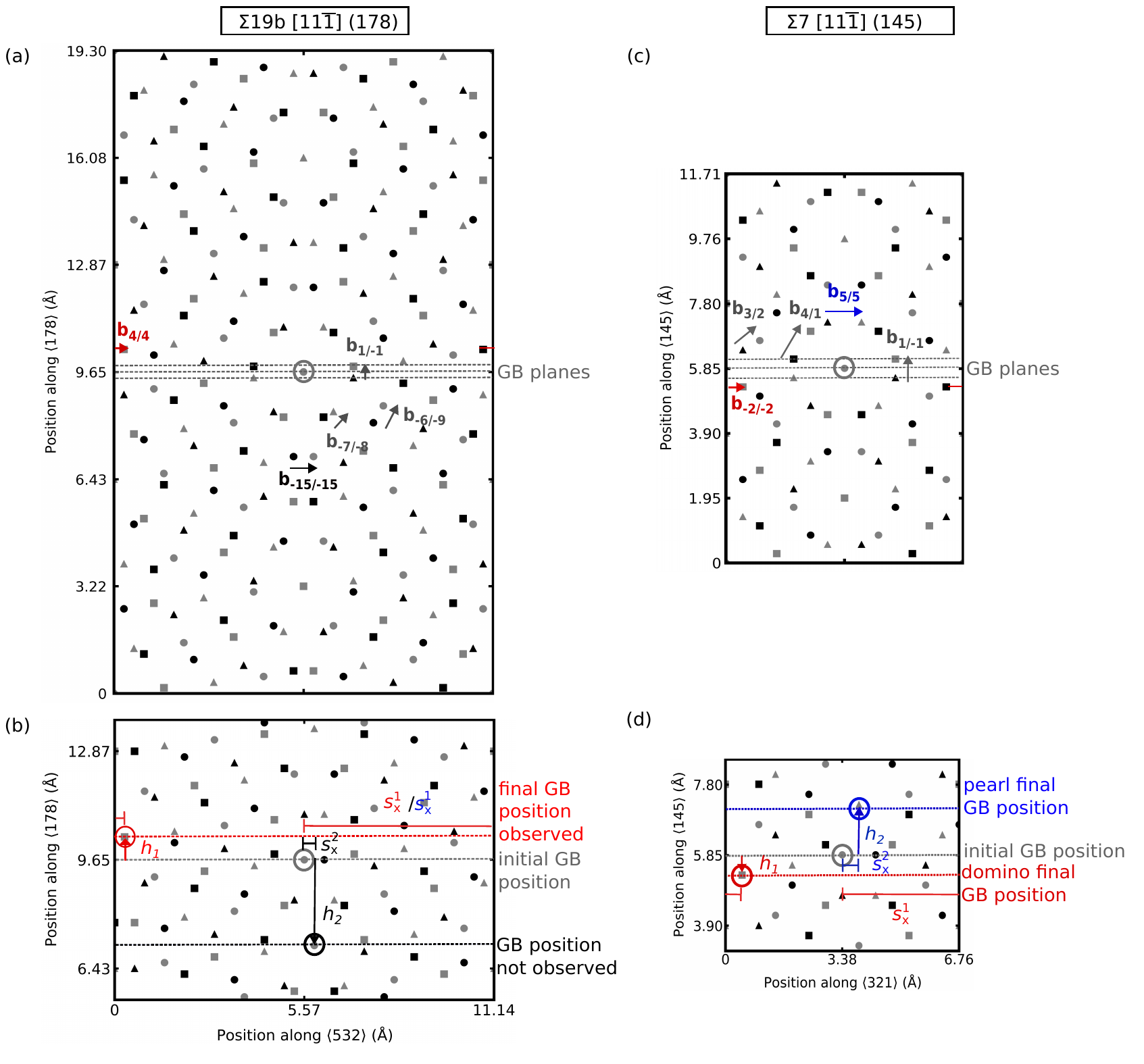}
\caption{\label{fig:dichromatic_pattern} Dichromatic patterns of the bicrystals. Gray and black symbols represent the atoms from the two different grains. Circle, triangle, and square indicate different layers along $z$ (ABC stacking sequence). (a) Dichromatic pattern of the $\Sigma$19b $[11\overline{1}]$ $(178)$ symmetric tilt GB in the $xy$ plane. Four DSC vectors $\mathbf{b}_{i/j}$ of low magnitude are marked by arrows, where $i$ and $j$ indicate the starting and ending plane as counted from the coincidence site. The equivalent DSC vectors $\mathbf{b}_{4/4} = \mathbf{b}_{-15/-15}$ are marked with red and black arrows and lie within the (178) GB plane. They are therefore the shortest possible Burgers vectors occurring during conservative shear-coupling (no climb). (b) Pattern after moving the top crystal by $\mathbf{b}_{4/4}$. If we imagine---without loss of generality---that the original GB plane passed through the coincidence site marked by a gray circle, the new GB plane must also pass through a coincidence site (red circle or black circle), because the GB structure must be equivalent before and after a disconnection has moved through the GB. Possible new GB planes therefore move in $\langle178\rangle$ direction by step heights of either $h_1$ (observed in both domino and pearl) or $h_2$ (not observed). At the same time, the new coincidence sites are also shifted along the $\langle532\rangle$ ($x$) direction. This shift can be interpreted as a shift of the atomic strucure of the GB to the left or right. For $h_1$, this shift $s_x^1$ corresponds to approximately half a unit cell, while for $h_2$ the shift $s_x^2$ is very small. (c)  Dichromatic pattern of the $\Sigma$7 $[11\overline{1}]$ $(145)$ symmetric tilt GB in the $xy$ plane. The DSC vectors $\mathbf{b}_{5/5} = \mathbf{b}_{-2/-2}$ (red and blue arrow) lie within the (145) GB plane and are thus the relevant Burgers vectors for shear coupling. (d) Pattern after moving the top crystal by $\mathbf{b}_{5/5}$. We similarly find the pairs $h_1$/$s_x^1$ and $h_2$/$s_x^2$ for possible new GB planes. For $\Sigma7$, however, both were observed: $h_1$/$s_x^1$ for domino and $h_2$/$s_x^2$ for pearl.}
\end{figure*}

It is known that line defects with dislocation character can exist on GBs \cite{Bollmann_1970,Ashby_1972,Hirth_1973, Hirth1996,Hirth_2006, Han2018}. These are called secondary GB dislocations or disconnections and possess both a Burgers vector $\mathbf{b}$ (leading to the dislocation character) and a step height $h$. Disconnections can only exist on and move along the GB, where they also introduce a step of the GB plane. Similar to a bulk dislocation, the structure of the GB on both sides of the disconnection is undisturbed. Thus, the Burgers vectors must be a displacement shift complete (DSC) vector. The DSC vectors can be obtained from the dichromatic pattern, i.e., by plotting the crystal lattices of both crystallites on top of each other (Fig.~\ref{fig:dichromatic_pattern}). The figure contains some example Burgers vectors, using the notation $\textbf{b}_{i/j}$ \cite{Serra1996}. The vector starts on a crystallographic plane $i$ parallel to the GB plane and ends on plane $j$. The integers $i$ and $j$ correspond to the distance of the plane from a coincidence site, i.e., vector $\textbf{b}_{-1/1}$ in Fig.~\ref{fig:dichromatic_pattern}(a) starts one plane below the coincidence site ($i = -1$) and ends one plane above ($j = 1$).

Since the glide plane of the active disconnections needs to be equal to the GB plane, the shortest possible Burgers vector for $\Sigma19$b is $\textbf{b}_{-15/-15} = \textbf{b}_{4/4} = [0.586,0,0]\,\text{\AA}$ [Fig.~\ref{fig:dichromatic_pattern}(a)]. The possible step heights are obtained by moving the top crystal by this shortest Burgers vector. The new resulting coincidence sites nearest to the original coincidence site position (gray circle) along $\langle178\rangle$ are marked with a red circle and a black circle, respectively [Fig.~\ref{fig:dichromatic_pattern}(b)]. The differences in the original and new coincident site positions along $\langle178\rangle$ result in step heights (GB migration distance) $h_1 = \SI{0.677}{\angstrom}$ ($4 \times h_0$)  and $h_2 = \SI{-2.535}{\angstrom}$ ($-15 \times h_0$), with unit step height $h_{0}= \frac{a/2}{|178|}$. The step height $h_1$ is significantly smaller than the next possible value, leading to the lowest step energy and thus the lowest $E_\text{core}$ out of all possible step heights. The only disconnection mode, which appears for both domino and pearl, is thus $(\mathbf{b},\textit{h})=([0.586,0,0]\,\text{\AA}, \SI{0.677}{\angstrom})$, resulting in $\beta = b_x/h = 0.865$ as obtained in the previous section. 

For $\Sigma7$, likewise, the shortest possible Burgers vector along the GB plane is $\textbf{b}_{-2/-2} = \textbf{b}_{5/5} = [0.966,0,0]\,\text{\AA}$ [Fig.~\ref{fig:dichromatic_pattern}(c)]. We obtain two step heights $h_1 = \SI{-0.557}{\angstrom}$ ($-2 \times h_0$) and $h_2 = \SI{1.395}{\angstrom}$ ($5 \times h_0$), with unit step height $h_{0}= \frac{a/2}{|145|}$ [indicated by red and blue circles in Fig.~\ref{fig:dichromatic_pattern}(d)]. This would mean that only the mode with $h_1$ should be active, having the lowest step energy.

\begin{figure}
    \includegraphics{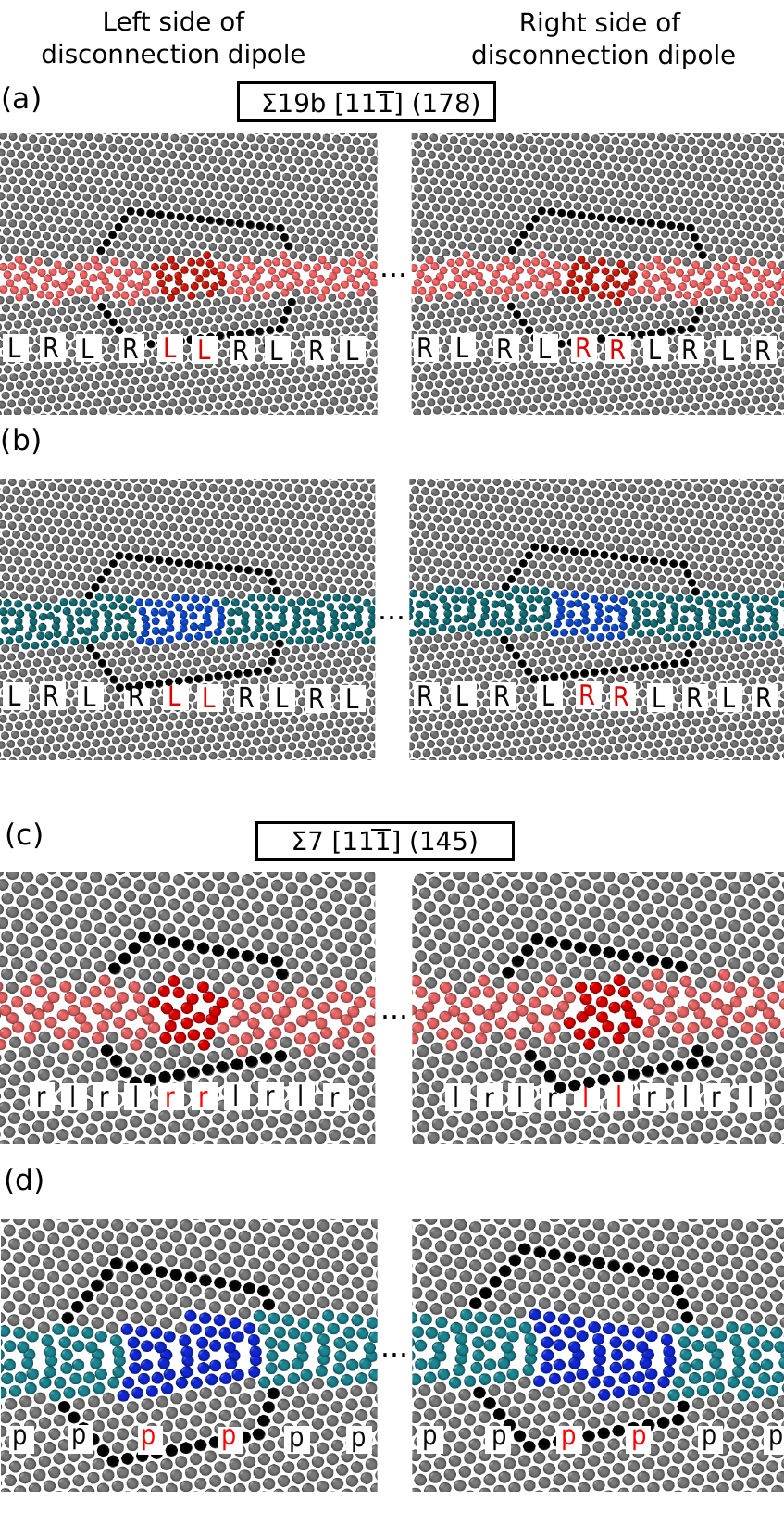}
    \caption{\label{fig:disconn_structures} Structure of the constructed disconnection dipoles for (a) domino and (b) pearl of $\Sigma$19b. On the left, the disconnection is $\mathbf{b} = [0.586,0,0]\,\text{\AA}$, $h=0.677$, with structural units \ldots{}LR\textbf{LL}RL\ldots{}. On the right, we have the opposite disconnection ($\mathbf{b} = [-0.586,0,0]\,\text{\AA}$, $h=-0.677$) with structural units \ldots{}RL\textbf{RR}LR\ldots{}. The disconnection core is therefore equivalent to a shift of the structural motifs by half a unit cell. (c) The same is observed for the domino complexion in $\Sigma$7 ($\mathbf{b} = \pm[0.966,0,0]\,\text{\AA}$, $h=\mp 0.557$). (d) For pearl, however, there is no such shift in the motifs, and the \ldots{}p\textbf{pp}p\ldots{} structural pattern is uninterrupted in the disconnection cores ($\mathbf{b} = \pm[0.966,0,0]\,\text{\AA}$, $h=\pm1.394$). The three dots between the images indicate that the dipole width $\delta$ is greater than shown here and the GB between the two disconnections is elided. Burgers circuits are drawn around the disconnections (black atoms) to verify the disconnection mode obtained after minimization.}
\end{figure}

\begin{figure}
\includegraphics{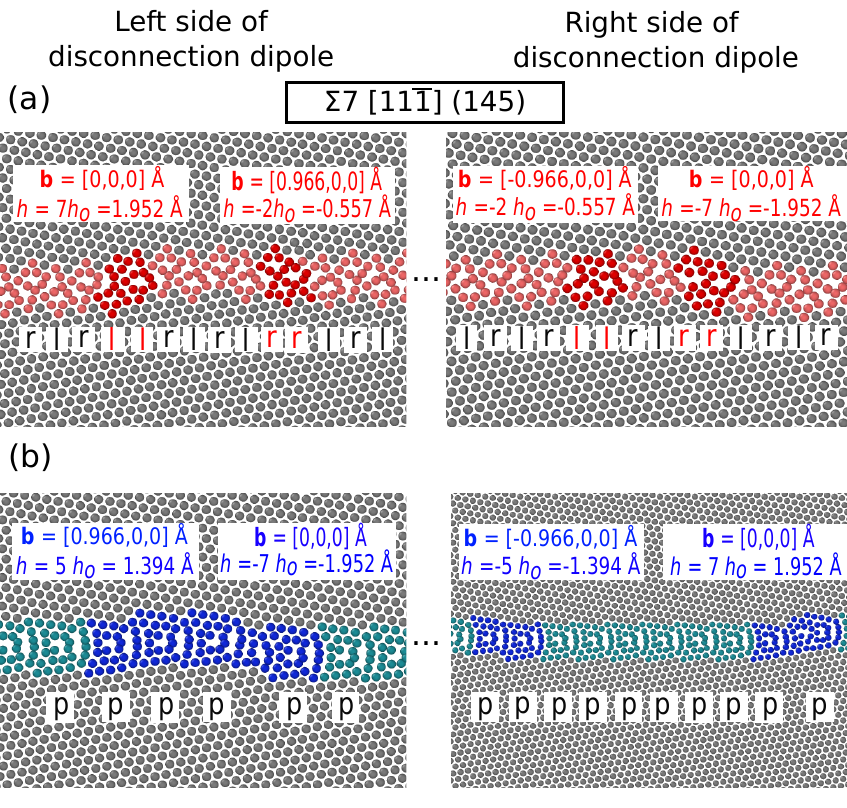}
\caption{\label{fig:S7nonactive} Construction of the non-active disconnection dipoles for $\Sigma$7 domino and pearl. When attempting to introduce the disconnection mode $(\mathbf{b}, h) = ([0.966, 0, 0]\,\text{\AA}, \SI{1.394}{\angstrom})$ for domino (a) and $([0.966, 0, 0]\,\text{\AA}, \SI{-0.557}{\angstrom})$ for pearl (b), the disconnections dissociate instead into the disconnections from Fig.~\ref{fig:disconn_structures}(c)--(d) and a step to compensate the difference towards the imposed, unfavorable step height. This indicates that the disconnection modes that were not observed during our shear-coupling simulations are not stable (even in molecular statics) due to their high core energy. Thus the structurally feasible disconnections in Fig.~\ref{fig:disconn_structures}(c)--(d) control the GB migration. }
\end{figure}

To investigate why the pearl complexion in $\Sigma7$ GBs nonetheless exhibits a step height of $h_2$, we constructed disconnection dipoles of all modes that were observed in the shear-coupling simulations and minimized the simulation cells with molecular statics [see Methods and Fig.~\ref{fig:GB_sketch}(c)]. This leads to a pair of opposite disconnections, namely $(\mathbf{b},h)$ and $(-\mathbf{b},-h)$ with separation $\delta$. Figure~\ref{fig:disconn_structures} shows the structure of the disconnection cores for $\Sigma19$b and $\Sigma7$ of the active modes. We verified that we obtained the desired defects by making a Burgers circuit around them \cite{Pond_1989, Rajabzadeh2014, Medlin2017} (black atoms, see Figs.~\ref*{fig:S19b_burgers_circuit} and \ref*{fig:S7_burgers_circuit} in the SM \cite{suppl} for details on the calculation of the Burgers vector and step height). When trying to construct the modes with step height $h_2$ for $\Sigma7$ domino and $h_1$ for $\Sigma7$ pearl (i.e., the modes that were not observed), these disconnections were unstable and dissociated into the favored disconnection mode plus an additional step (Fig.~\ref{fig:S7nonactive}). The active disconnection modes thus have a lower core energy and favorable core structure, compensating for the higher step height $h_2$ that occurs for the pearl complexion. As a result, domino has the mode $(\mathbf{b},\textit{h})=([0.966,0,0]\,\text{\AA}, \SI{-0.557}{\angstrom})$, resulting in $\beta = b_x/h = -1.734$, while pearl has the mode $([0.966,0,0]\,\text{\AA}, \SI{1.395}{\angstrom})$, resulting in $\beta = 0.692$, as also observed in the previous section.

Why are these specific core structures energetically favorable and why is there no difference in $\Sigma19$b? An answer can be found by referring back to the dichromatic patterns in Fig.~\ref{fig:dichromatic_pattern}. The vector between two equivalent positions in the dichromatic pattern before and after applying a Burgers vector \textbf{b} possesses both a vertical ($h$) and a horizontal component ($s_x$). The former is of course the step height, but the latter also has a physical meaning and is not identical to the Burgers vector \textbf{b}. The component $s_x$ determines how the atomic motifs on the GB shift along the $x$ direction, i.e., along the GB. For the active modes in $\Sigma19$b and $\Sigma7$ domino, $s_x$ corresponds to approximately half the unit cell. These structures also contain two motifs in our nomenclature (L/R and l/r). The disconnection core thus, e.g., results in a shift from \ldots{}lr\textbf{lr}lr\ldots{} to \ldots{}lr\textbf{ll}rl\ldots{} (cf.~Fig.~\ref{fig:disconn_structures}). When forcing the $\Sigma7$ domino complexion to assume the other step height, which is connected to $s_x \sim 0$, it would have to retain the order \ldots{}lr\textbf{lr}lr\ldots{} for its motifs at the disconnection core. Figure~\ref{fig:S7nonactive}(a) shows that this seems to be energetically and structurally very unfavorable, so that the lr defect dissociates into an ll and an rr defect. In $\Sigma7$ pearl, on the other hand, we only marked a single structural motif p in our unit cell. We can see from Figs.~\ref{fig:disconn_structures}(d) and \ref{fig:S7nonactive}(b) that pearl indeed prefers the \ldots{}p\textbf{pp}p\ldots{} ordering and that splitting the p-motif would be connected to an energy cost. Thus pearl migrates with the mode $h_2$, where $s_x^2 \sim 0$, which leads to a low-energy disconnection core structure.

These results show that not all crystallographically possible disconnection modes actually exist. Indeed, for the same macroscopic GB parameters and for the same boundary conditions of our shear-coupling simulations, the atomic structure of both the complexions as well as the disconnection cores determine which of these modes will be active in a given complexion.

\subsection{\label{sec:coupling}Critical stress of GB migration}

\begin{figure}
\includegraphics{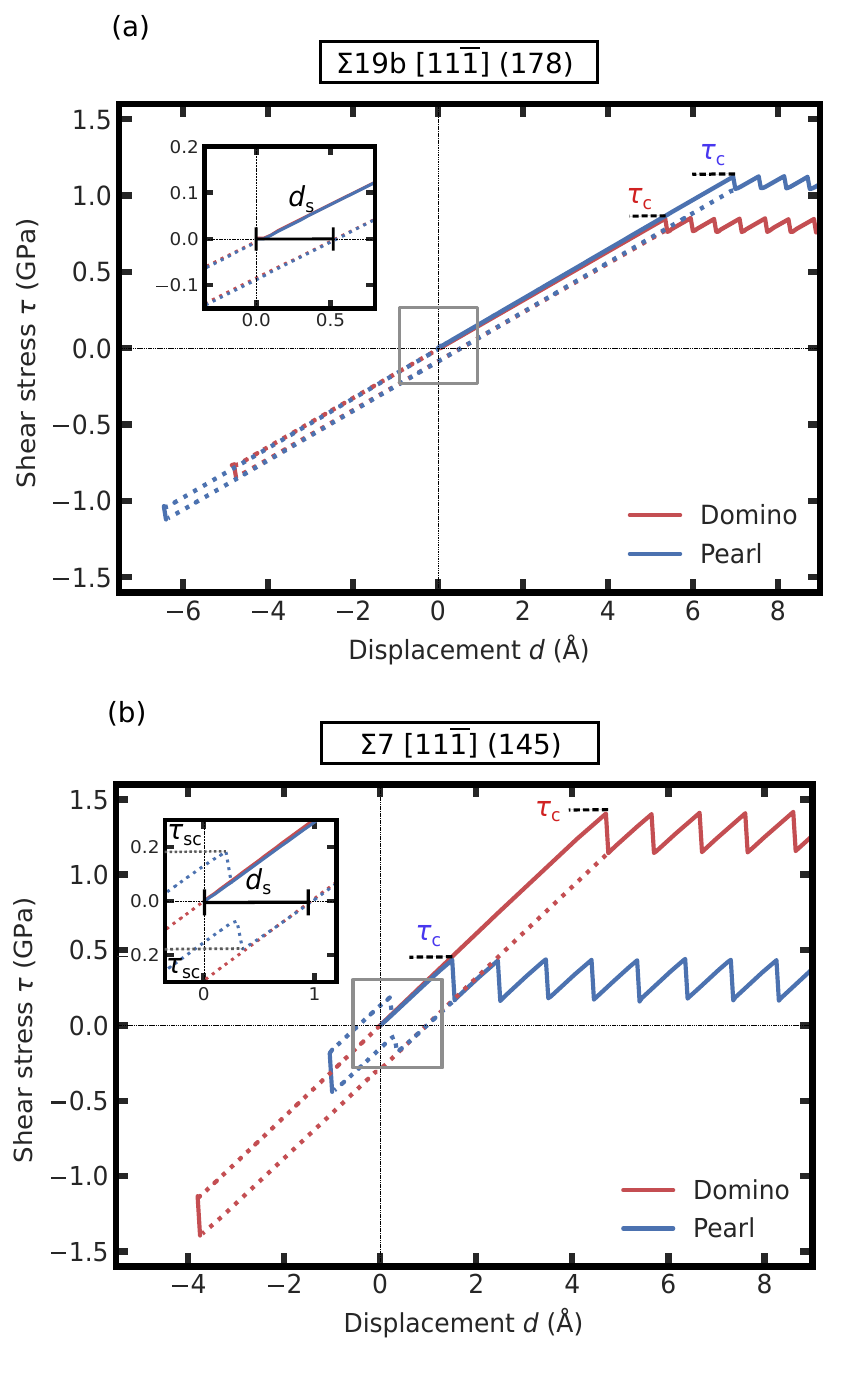}
\caption{\label{fig:shear_stress} (a) Shear stress response to displacement at 0 K plotted for domino and pearl of $\Sigma$19b GBs. Each shear stress drop observed corresponds to a unit step GB migration. The maximum shear stress is the critical shear stress $\tau_{c}$ required for GB migration. In inset, the residual displacement $d_s \approx \SI{0.550}{\angstrom}$ is the shear obtained for GB migration by a unit step for both domino and pearl. The migration normal to the GB plane from the initial position obtained from atomistic simulations is \SI{0.647}{\angstrom} and \SI{0.686}{\angstrom} for domino and pearl, respectively. (b) The same for $\Sigma$7 GBs, with $d_s \approx \SI{1}{\angstrom}$ and migration distances of \SI{-0.490}{\angstrom} (domino) and \SI{1.401}{\angstrom} (pearl). When switching the displacement direction, a transition between degenerate microstates of the pearl complexion occurs at $\tau_\text{sc}$ (see inset).}
\end{figure}

While the $\Sigma19$b complexions exhibit the same shear-coupling factor, it is nevertheless of interest if the critical shear stress required for GB migration differs. In molecular statics simulations, displacement was imposed and the GB migration velocity $v_\perp$ only depends on the applied shear velocity $v_\parallel$ and $\beta$. In reality, it is often the case that a given stress is applied, so a relevant figure of merit is the critical stress $\tau_c$ required to start GB migration. This can be obtained by monitoring the reaction forces at the boundary where the shear is applied. We started with deformation in molecular statics ($T = \SI{0}{K}$). As the displacement $d$ increases, the shear stress $\tau$ increases linearly in the elastic regime until stress drops occur (Fig.~\ref{fig:shear_stress}). This saw-tooth behavior is similar to earlier reported simulation studies \cite{Cahn2006, Mishin2007, Ivanov2008,Rajabzadeh2013}. We note that the residual displacement $d_s$ after unloading at the first stress drop (dotted lines) together with the GB migration distance $h$ after a single migration event \footnote{The GB migration distance $h$ traveled by the GB during a single event is obtained from atomistic simulations by marking an equivalent position in the structural unit (e.g., in \textbf{R}, \textbf{r}, or \textbf{p}) before and after the migration step. The difference in their atomic positions along the $y$ direction is then equal to $h$.} can also be used to calculate $\beta = d_s / h$. We obtain the same values as earlier within the expected error bounds. Continuing the unloading simulations to negative displacements until the first stress drop and then reversing the loading direction again leads to the expected hysteresis. The results for positive and negative displacement directions are symmetric.

For $\Sigma$19b only a single disconnection mode was observed in all cases. However, we obtained different critical shear stresses $\tau_c$ for pearl (\SI{1.117}{GPa}) and domino (\SI{0.849}{GPa}), meaning that the complexions affect the activation barrier of the shear-coupled motion. In contrast, the differences in critical shear stress of the $\Sigma7$ complexions are expected due to the different disconnection modes. We find $\tau_c = \SI{0.434}{GPa}$ for pearl and \SI{1.405} {GPa} for domino. We also find that there is an additional, small stress drop when reversing the displacement direction for the pearl complexion (critical shear stress $\tau_\text{sc} = \SI{0.18}{GPa}$). The reason is that two degenerate states of the pearl complexion exist \cite{Brink_2023}. These are slightly sheared either to the left or to the right \cite{Brink_2023}. Their degeneracy disappears under applied shear stress, meaning that depending on the direction of the applied displacement, either one or the other state is stabilized. This only affects simulations where the displacement direction is switched and does not influence the shear-coupled motion otherwise, which is why we do not further discuss this effect.

\begin{figure}
\includegraphics{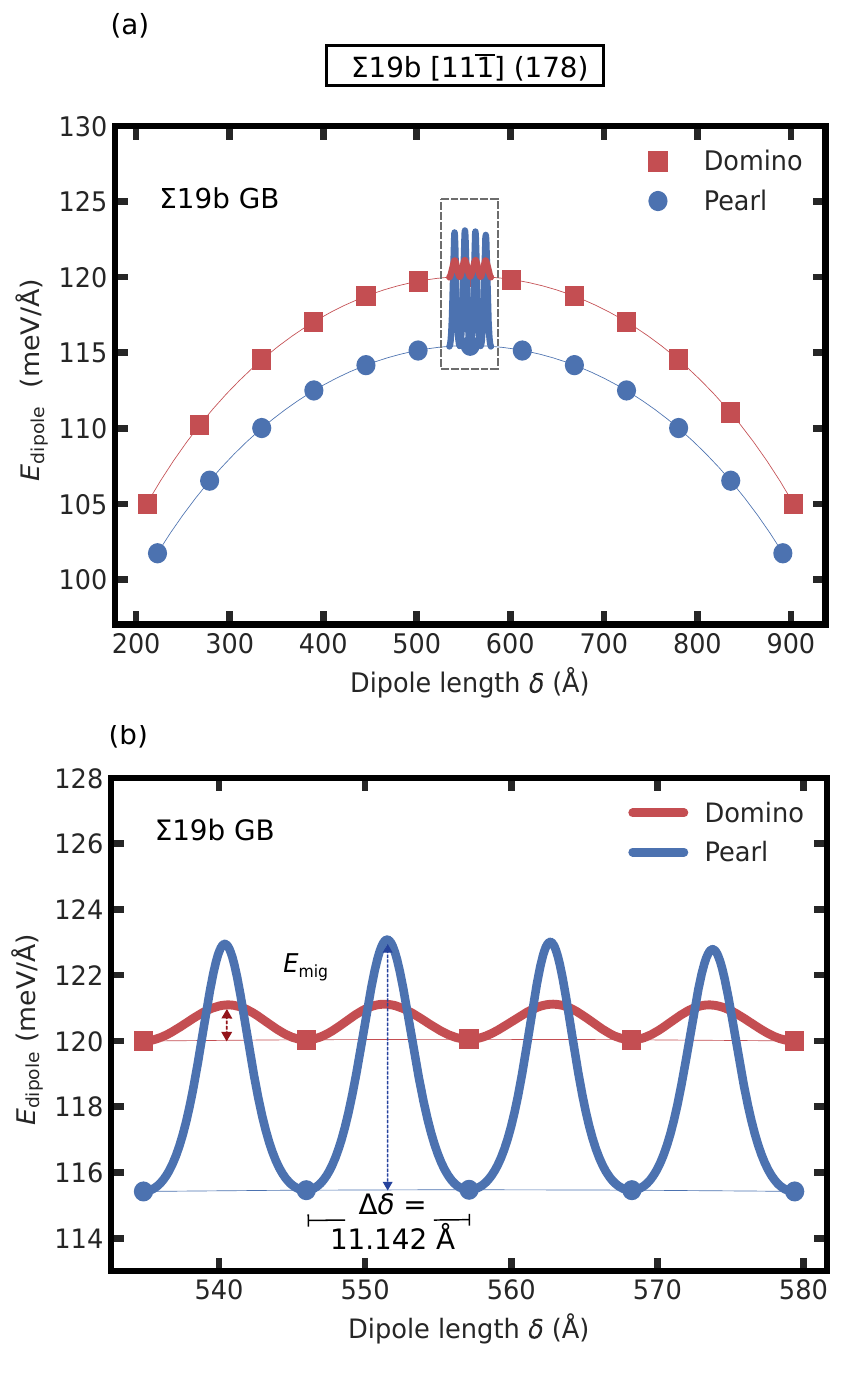}
\caption{\label{fig:energy} (a) The formation energies $E_\text{dipole}$ of disconnections for different dipole widths $\delta$ of the disconnection mode $(\mathbf{b}, h) = ([0.586, 0, 0]\,\si{\angstrom}, \SI{0.677}{\angstrom})$ are plotted for domino and pearl complexions of $\Sigma$19b GBs. (b) Zoom of the gray area in (a). The energies of dipoles during their migration (thick lines) were obtained by NEB between the indicated local minima (squares and circles). The difference in disconnection widths of local minima taken for NEB study is given by $\Delta \delta$. The disconnection migration barrier $E_\text{mig}$ is highlighted by red and blue arrows for domino and pearl.}
\end{figure}

\begin{figure}
\includegraphics{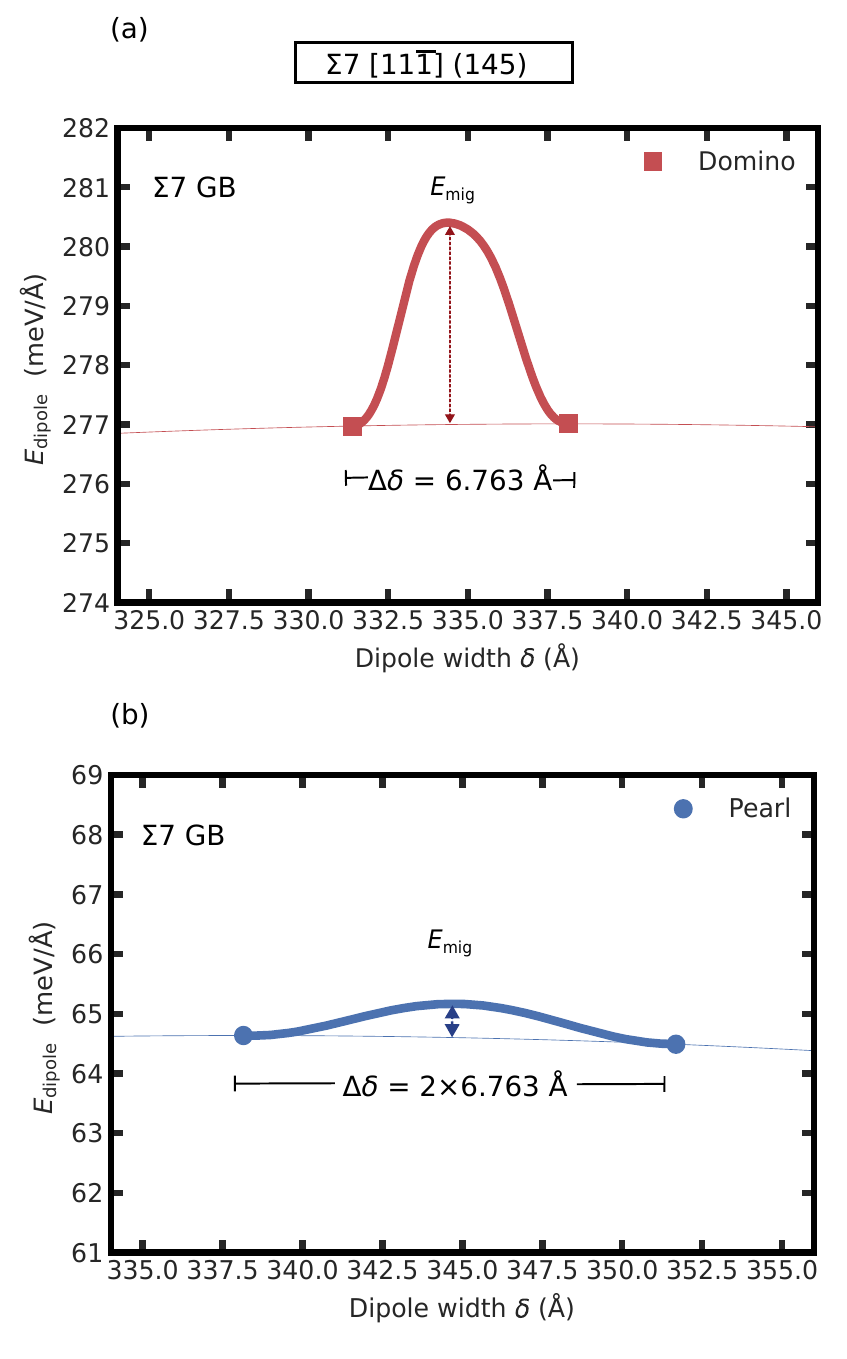}
\caption{\label{fig:S7_energy}The formation energies $E_\text{dipole}$ of disconnections in $\Sigma$7 GBs for two different dipole widths $\delta$  are plotted for (a) domino with the disconnection mode $(\mathbf{b}, h) = ([0.966, 0, 0]\,\si{\angstrom}, \SI{-0.557}{\angstrom})$ and (b) pearl with the disconnection mode $([0.966, 0, 0]\,\si{\angstrom}, \SI{1.394}{\angstrom})$.  The energies of dipoles during their migration (thick lines) were obtained by NEB between the indicated local minima (squares and circles). The difference in disconnection widths of local minima taken for NEB study is given by $\Delta \delta$. The disconnection migration barrier $E_\text{mig}$ is highlighted by red and blue arrows for domino and pearl. }
\end{figure}

In order to understand the differences in the critical shear stress, we first obtained an estimate of the relative formation energies of the disconnection dipoles. We use the differences in dipole energy as discussed in Sec.~\ref{sec:method}, Eq.~\ref{eq:E_delta}, for domino and pearl as an indicator of this. The dipole energy depends on the system size $L_x$ and dipole width $\delta$, wherefore we constructed several dipoles with different $\delta$ as described above. (An exploration of the influence of $L_x$ is provided in Fig.~\ref*{fig:Emax} in the SM \cite{suppl}. The results follow the expectations from the theory in Eq.~\ref{eq:E_max}.). Note that this is effectively a 2D model of dipole formation as parallel disconnection lines instead of disconnection loops \cite{Han2018}. We found that the critical shear stress is not affected by the thickness of our samples up to the maximum investigated thickness of \SI{12}{nm} (Fig.~\ref*{fig:shearstressT} in the SM \cite{suppl}), which means that the present samples are still thin enough that the 2D model applies to our simulations. We will discuss the implications for experimental samples and thick systems later, but for now use the present results at the very least as a qualitative indicator.

\begin{table}
\caption{\label{tab:fitparameters}Disconnection dipole parameters obtained from fitting Eq.~(\ref{eq:E_delta}) to \textit{E}\textsubscript{dipole} from the simulations.
We list the system size $L_y$, the parameter $K$ describing anisotropic crystal elasticity, the effective disconnection core size $\delta_0$, the dipole energy $E^*_\text{dipole}$ for $\delta = L_x/2$, and the Peierls barrier $E_\text{mig}$, each for the domino and pearl complexions. The dipole energy differences $\Delta E^*_\text{dipole} = E^\text{* domino}_\text{dipole} - E^\text{* pearl}_\text{dipole}$ are also listed.}
\begin{ruledtabular}
\begin{tabular}{lcc}
CSL type &  $\Sigma$7 & $\Sigma$19b\\ 
 \hline
  $L_{y}$ ({\AA}) & 1171.61 & 1929.540  \\[2pt]
$K_\text{domino}$ (meV/{\AA}$^{3}$) & 79.214 &77.462  \\[2pt]
$\delta_0^\text{domino}$ ({\AA}) & 5.082 &3.540  \\[2pt]
$E_\text{dipole}^\text{* domino}$ (meV/{\AA}) & 277.015 & 120.027 \\[2pt]
$E_\text{mig}^\text{domino}$ (meV/{\AA}) & 3.430 & 1.083\\[2pt]
 \hline  
 $L_{y}$ ({\AA}) & 1170.29 & 1929.540  \\[2pt]
$K_\text{pearl}$ (meV/{\AA}$^{3}$) & 79.214 & 76.822   \\[2pt]
$\delta_0^\text{pearl}$ ({\AA}) & 92.652  & 4.067  \\[2pt]
$E_\text{dipole}^\text{* pearl}$ (meV/{\AA}) & 64.634 & 115.460 \\ [2pt]
$E_\text{mig}^\text{pearl}$ (meV/{\AA}) & 0.532 & 7.618 \\ [2pt]
 \hline
 $\Delta E^*_\text{dipole}$ (meV/{\AA}) & 212.381 & 4.567  \\
\end{tabular}
\end{ruledtabular}
\end{table}

The data points (circles and squares) in Fig.~\ref{fig:energy}(a) show the results for $\Sigma19$b. While the dipole energy depends on $\delta$, the difference between domino and pearl is approximately constant and disconnection dipoles in domino complexions have a consistently higher energy. This contradicts our findings that the critical stress to move $\Sigma19$b pearl GBs is higher. In contrast, the data for $\Sigma7$ GBs in Fig.~\ref{fig:S7_energy} is in accordance with our earlier results: The dipole energy and critical shear stress for domino complexions is much higher than for pearl complexions. The lines connecting the data points represent fits of Eq.~\ref{eq:E_delta}, parameters are listed in Table~\ref{tab:fitparameters}. The resulting parameters are comparable to those in previous studies, see Appendix~\ref{sec:GBcomparison} for details. We also note that the parameter $K$, which represents the anisotropic elastic response of the material, can be calculated analytically from the stiffness tensor of the crystals \cite{Eshelby_1953,Foreman_1955,Stroh_1958, Hirth1983}. Following the methods of Eshelby et al.~\cite{Eshelby_1953} and Stroh~\cite{Stroh_1958}, we obtained $K = \SI{76.7}{meV/\angstrom^3}$ for both $\Sigma$19b and $\Sigma$7 GBs, in rough accordance with the fitted values.

Apart from the dipole formation, disconnections also need to move in order to facilitate GB migration. In analogy to bulk dislocations, there is also a Peierls barrier $E_\text{mig}$ for disconnections, requiring a critical Peierls--Nabarro stress $\tau_\text{mig}$ to move \cite{Peierls1940, Nabarro1947, Nabarro1952, Khater2012, MacKain2017,Chen_2020}. The atomic configuration at the saddle point of this barrier is not stable and can therefore not be explored with simple molecular statics calculations. The minimum energy path for the migration of a disconnection along the GB, extending the dipole width by the distance $\Delta\delta$ between two minima, was calculated using NEB (thick lines in Figs.~\ref{fig:energy} and \ref{fig:S7_energy}). For $\Sigma19$b we find that $E_\text{mig}$ is much higher for pearl (\SI{7.6}{meV/\angstrom}) as compared to domino (\SI{1.1}{meV/\angstrom}), leading to a steeper energy landscape during disconnection migration and a high $\tau_\text{mig}$. For $\Sigma7$ GBs, $E_\text{mig}$ is higher for domino (\SI{3.430}{meV/\angstrom}) than for pearl (\SI{0.532}{meV/\angstrom}). This means that the difference between critical shear stresses in $\Sigma19$b GBs is dominated by the Peierls barrier, at least at low temperatures. Both the formation and Peierls barrier of domino in $\Sigma7$ GBs are much higher than for pearl, making this case much more straightforward.

We could not find conclusive structural reasons for the large differences in $\Sigma19$b GBs, where all complexions share the same active disconnection mode. An investigation of atomic displacements during GB migration did not reveal any obvious differences between pearl and domino that could be responsible for higher migration barriers (Appendix~\ref{sec:atom_shuffling}). A possible explanation could be the generally higher GB excess volume of domino compared to pearl \cite{Brink_2023}: With more available free volume, the barriers for atomic rearrangement might be lower. However, due to the limited available data, this remains a hypothesis.

These simulations are of course probing an idealized case. In reality, in thicker systems, GB migration proceeds through the formation of islands of atoms or disconnection loops \cite{Race2015, Hadian2016, Han2018}, most likely nucleated heterogeneously at defects \cite{Hadian2016}. The results of Sec.~\ref{sec:bicrystallography} should not be affected by this, since they are intrinsic to the structure of the complexions and their crystallography and not due to the exact nucleation process. The energy barriers and critical stresses calculated in the present section, on the other hand, represent theoretical values. We nevertheless argue that they are useful because they can give qualitative insights into the influence of complexions on GB properties, such as GB migration.

\subsection{\label{sec:tempdiscmigration}Temperature effects}

\begin{figure}%
\includegraphics{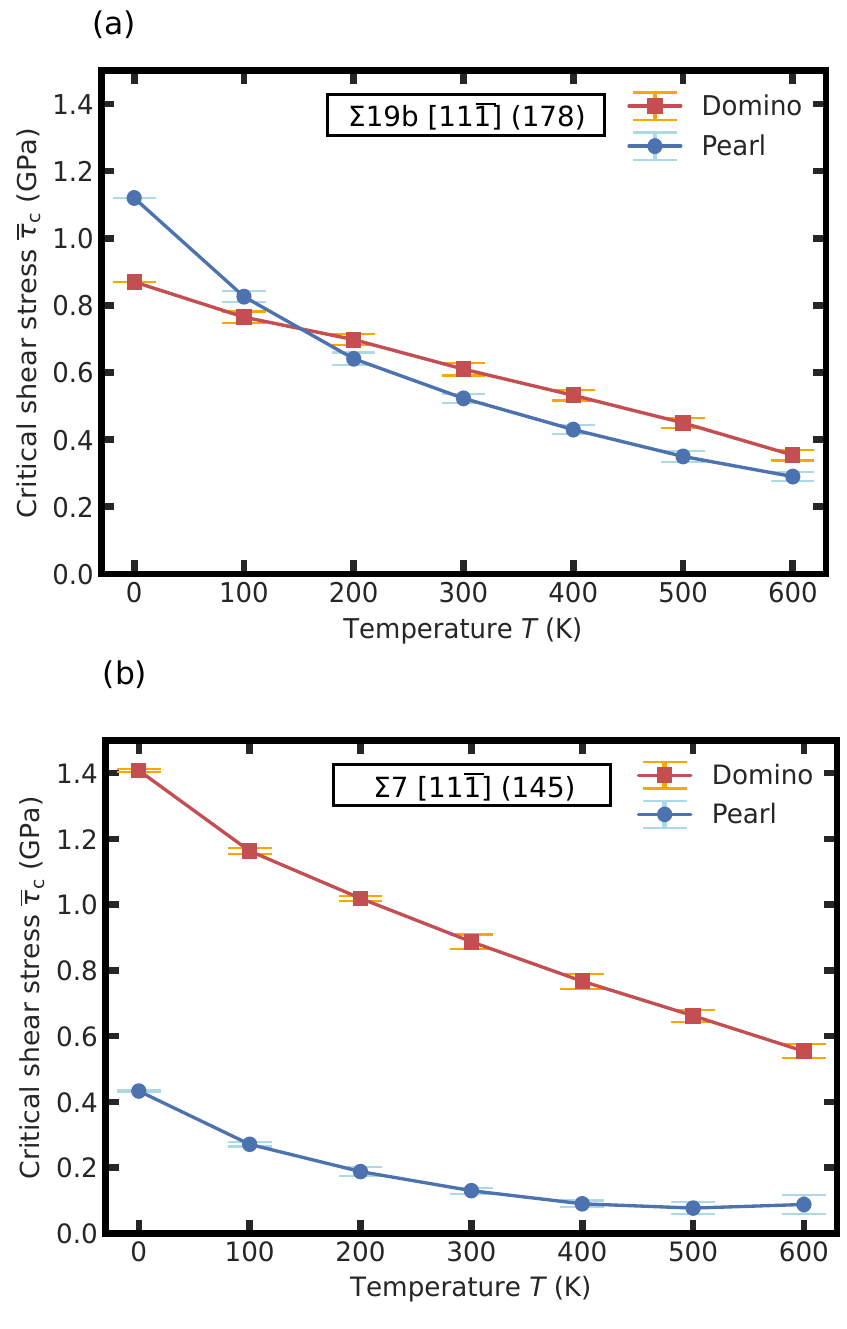}
\caption{\label{fig:T_shear}  The critical shear stress $\tau_c$ for GB migration as a function of temperature. (a) For $\Sigma$19b, at temperatures of around \SI{200}{K} and above, the critical stress $\tau_c$ of pearl drops below the one for domino. (b) For  $\Sigma$7, the critical stress $\tau_c$ of pearl migration is always below domino.}
\end{figure}

\begin{figure*}%
\includegraphics{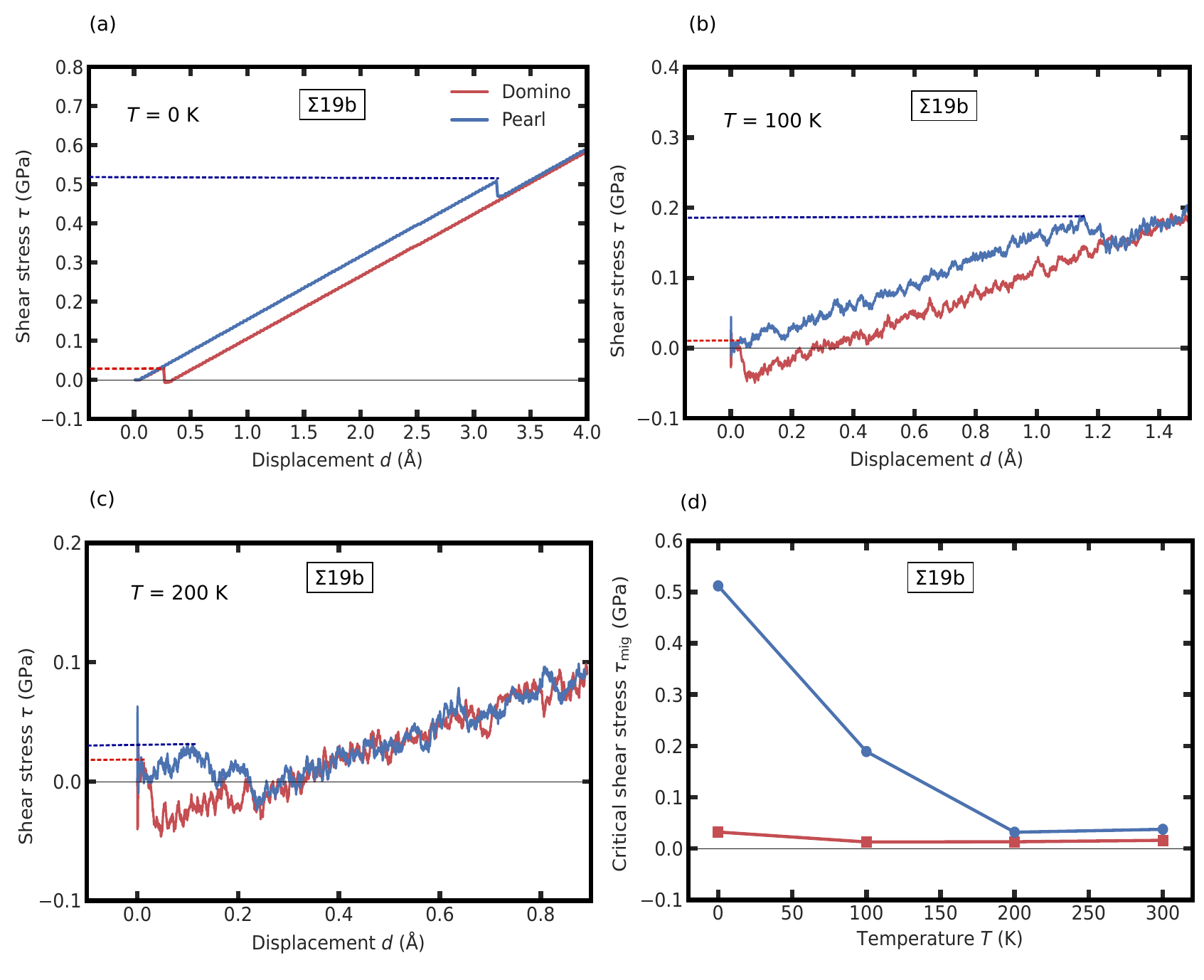}
\caption{\label{fig:disconn_shearstress} (a)--(c) Shear stress--displacement curves of $\Sigma$19b $[11\overline{1}]$ $(178)$ GBs with pre-existing disconnection dipoles of width $\delta^* = L_x/2$. The shear stress response to displacement at temperatures (a) 0\,K, (b) 100\,K, and (c) 200\,K is plotted. The critical stress $\tau_\text{mig}$ for disconnection migration is indicated by the horizontal dotted lines. (d) It reduces with increasing temperature, dropping to close to zero at \SI{200} {K} and above for both domino and pearl.}
\end{figure*}

In addition to the molecular statics simulations, we also repeated the analysis of shear coupling factor and critical stress at finite temperatures with MD simulations. We found the shear coupling factor to be independent of both applied shear velocity and temperature (Fig.~\ref*{fig:velocity_beta_temp} in the SM \cite{suppl} shows this for the example of $\Sigma$19b). For temperatures up to $T = \SI{600}{K}$, we averaged the critical stress over several migration events (which are each equal to a stress drop), each time recording the maximum of the stress curve. At \SI{700}{K} and above, the $\Sigma19$b domino phase transitions into pearl within the first few migration events, making it impossible to extract meaningful data. (The domino complexion is metastable over the whole temperature range \cite{Meiners2020}, but the higher temperatures accelerate its transition to pearl even within the short simulation timescale). We thus restricted all simulations to the lower temperatures.

The results for $\Sigma19$b are shown in Fig.~\ref{fig:T_shear}(a). Interestingly, the critical stress of pearl is only higher than that of domino up to a point between \SI{100}{K} and \SI{200}{K}. If we compare the Peierls barrier $E_\text{mig}$ to the absolute formation energy $E_\text{dipole}$, however, we notice that the former is quite small (Fig.~\ref{fig:energy}). Only due to its steepness, is it connected to a high stress. It is conceivable that the thermal energy would be sufficient to help overcome this small barrier, so that in the end only the formation energy matters. We tested this by starting with systems that already have a disconnection dipole of width $\delta^* = L_x/2$ inserted before applying shear. We can thus probe only $\tau_\text{mig}$, the critical stress for disconnection migration. Figure~\ref{fig:disconn_shearstress}(a) shows that domino has $\tau_\text{mig} \approx \SI{0}{GPa}$, while pearl has $\tau_\text{mig} \approx \SI{0.5}{GPa}$  for $\Sigma19$b. The difference in critical shear stress $\tau_c$ for GB migration (Fig.~\ref{fig:shear_stress}) is roughly \SI{0.27}{GPa}, which is smaller. That is not surprising, since there is also a higher stress connected to the nucleation of domino. With increasing $T$, however, the disconnection migration barrier can be overcome more and more easily, resulting in $\tau_\text{mig} \approx \SI{0}{GPa}$ for both domino and pearl already at $T = \SI{200}{K}$ [Fig.~\ref{fig:disconn_shearstress}(b)--(d)]. This can explain the temperature dependence: It is easier to nucleate disconnections in the pearl complexion, but at low temperatures these disconnections have to cross high barriers to move. These barriers, however, are only high compared to domino and can be overcome with thermal energy. At room temperature and above, pearl GBs are easier to migrate since the GB migration is limited by defect nucleation.

For $\Sigma$7, the critical stresses are shown in Fig.~\ref{fig:T_shear}(b). Contrary to $\Sigma$19b, the temperature dependence of the critical stress is straightforward to understand: Domino complexions always have higher formation and Peierls barriers for disconnection dipoles than pearl and thus exhibit a higher $\tau_c$. The obtained values are in all cases of similar magnitude as other GBs in fcc metals (see Appendix~\ref{sec:GBcomparison}).

\section{\label{sec:conclusion}Conclusion}

We investigated elementary mechanisms of shear-coupled GB motion of two complexions, namely, domino and pearl in Cu $[11\overline{1}]$ tilt GBs. It is known from previous literature that several disconnection modes can exist for the same GB, leading, e.g., to opposite migration directions under the same applied shear. Which mode is active, however, could not be predicted. In this work, we show that the selection of the disconnection mode, identified by a Burgers vector and step height, depends on the complexion. In $\Sigma$19b GBs, the pearl and domino complexions exhibit the same modes, while in $\Sigma$7 GBs these complexions migrate in opposite directions. The selection of the active mode is a result of the structural motifs of the complexions, which dictate the core structures of the disconnections. We found that the combination of structural GB motifs in the disconnection core lead to significantly different core structures, some of which are energetically unstable, thereby selecting the active mode as the energetically favorable core structure. Even if the complexions exhibit the same mode (here in the $\Sigma$19b GBs), their different atomic structures also affect the critical shear stress required to move the GBs, at least in our model setup with symmetric, defect-free GBs.

\begin{acknowledgments}
The authors thank Nicolas Combe and Anupam Neogi for helpful discussions. This project has received funding from the European Research Council (ERC) under the European Union’s Horizon 2020 research and innovation programme (Grant agreement No. 787446; GB-CORRELATE). 
\end{acknowledgments}

\appendix

\section{\label{sec:shear_coupling_complexions}Shear coupling factor of complexions}

\begin{figure*}
\includegraphics{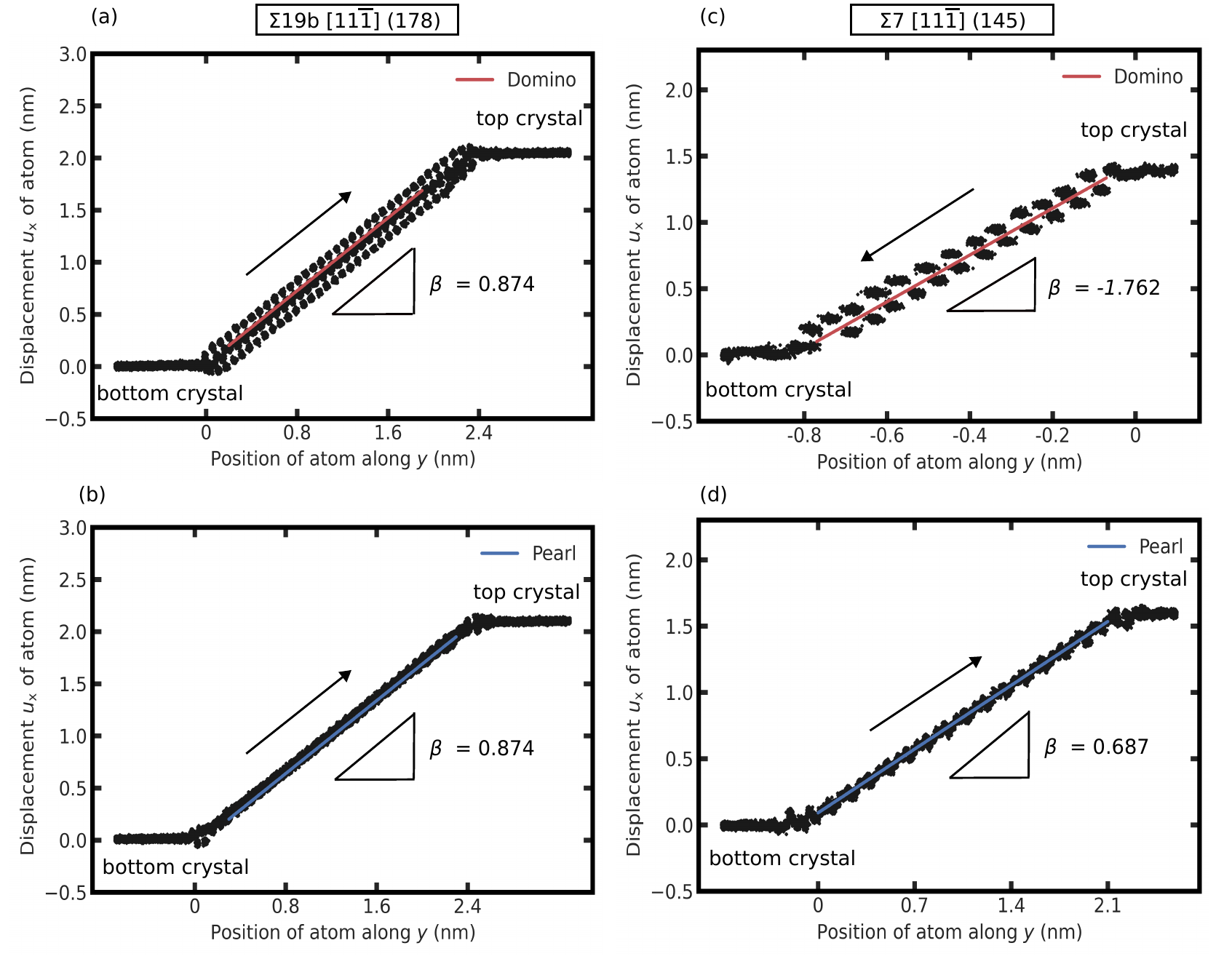}
\caption{\label{fig:beta_slope} Atomic displacements $u_x$ as a function of the atom's position along the $y$ axis. Note that $y=0$ corresponds to the initial GB position. The shear coupling factor $\beta$ is the slope (red and blue lines) of these graphs and was obtained by linear regression (Eq.~\ref{eq:beta}). Result for simulations at $\SI{300}{K}$ for $\Sigma$19b GBs after $\SI{20}{ns}$ with (a) domino and (b) pearl complexions and for $\Sigma7$ GBs after \SI{16}{ns} with (c) domino and (d) pearl complexions are shown. The regions belonging to the bottom crystal (left part of the graph with zero displacement) and the regions belonging to the top crystal (right part of the graph with $u_x = d$) are strain-free (constant displacement), while the region traversed by the GB was sheared. The arrows indicate the direction of GB migration; note that for domino in $\Sigma7$ (c), the bottom crystal shrinks, while it grows in all other cases (a)--(b), (d).}
\end{figure*}

To more accurately calculate $\beta$, we can record the displacement $u_x$ of the atoms and plot it as a function of the atomic position normal to the GB (Fig.~\ref{fig:beta_slope}). The slope can be obtained by linear regression and corresponds directly to $\beta$. Here, the shear coupling factor is obtained for simulations at $\SI{300}{K}$ after $\SI{20}{ns}$ in $\Sigma$19b GBs, shown in Fig.~\ref{fig:shearcoupling}(a)--(b), and after $\SI{16}{ns}$ in $\Sigma$7 GBs, shown in Fig.~\ref{fig:shearcoupling}(c)--(d).

\begin{figure}
\includegraphics{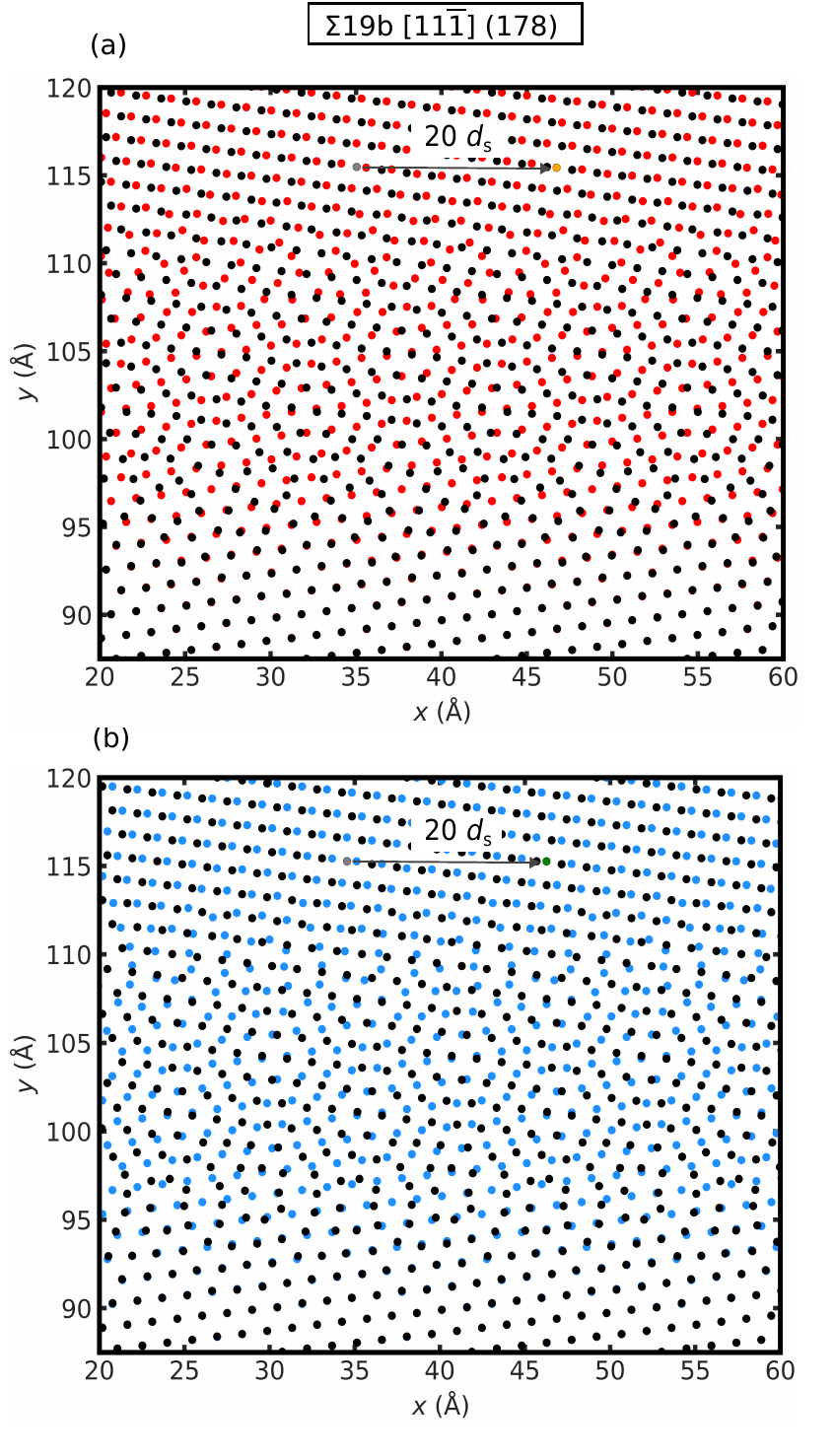}
\caption{\label{fig:atoms_shuffle} Atom positions in (a) domino and (b) pearl of $\Sigma$19b GBs before (black) and after (red/blue) shear-coupled GB motion. These images are after $n = 20$ unit steps of GB migration, which correspond to a GB migration distance of $n\mathit{h} = \SI{13.540}{\angstrom}$ along $y$ and a shear displacement $n|\mathbf{b}| = 20 d_s = \SI{11.720}{\angstrom}$ along $x$. In the traversed region, an image similar to the dichromatic pattern appears due to overlaying atoms from before the migration, which belong to the top crystal, and after the migration, which now belong to the bottom crystal. An additional offset between the atoms in the pattern is due to the microscopic degrees of freedom, i.e., the top and bottom crystal are shifted against each other depending on the complexion. Furthermore, the pattern at the start and end of the migration region is somewhat smeared out, indicating that the atomic jumps during the GB migration do not necessarily go from the initial to the final position, but can also occupy intermediary positions.}
\end{figure}

\begin{figure}
\includegraphics{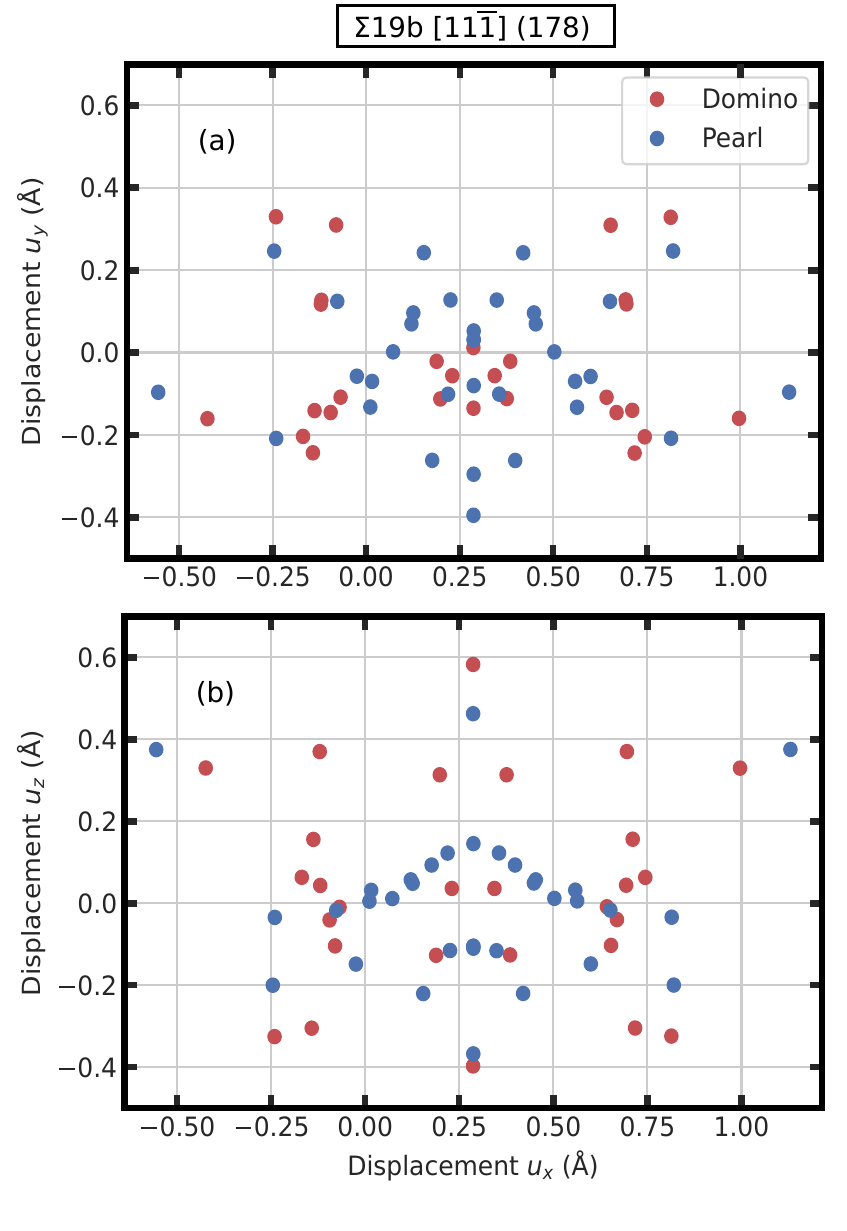}
\caption{\label{fig:dispmag} Displacements of the atoms in the GB during a single GB migration step in $\Sigma$19b GBs, plotted by (a) $u_x$ and $u_y$ components and (b) $u_x$ and $u_z$ components. Here, we define GB atoms as those atoms that were not identified as fcc either before or after the migration event by the polyhedral template matching method  \cite{Larsen_2016} in OVITO \cite{Stukowski2009}. The displacements are symmetric around $|\mathbf{b}|/2 = \SI{0.293}{\angstrom}$ in $x$ direction, but do not correspond to DSC vectors. This indicates that the jumps during one migration step go to intermediate positions, before arriving at their final environment in the defect-free crystal after several migration steps.}
\end{figure}

\section{\label{sec:atom_shuffling}Atomic shuffling during GB migration}

The simulations of perfect complexions go through a cycle of nucleating, propagating and annihilating the disconnections. For  $\Sigma$19b, we seem to observe the same disconnection mode for domino and pearl (same $\beta$), the differences in shear stress required to move the disconnection can also be due to the differences in the atomic shuffling during GB migration for both the complexions \cite{Chesser_2021,Chesser_2022}. The atomic shuffling during GB migration for both the complexions was observed after 20 shear stress drops in the molecular statics calculations (corresponding to $n = 20$ unit steps of GB migration). The mean GB plane moved a total distance of $n\mathit{h} = \SI{13.540}{\angstrom}$ along $y$ and the shear displacement $20 d_s = n|\mathbf{b}| = \SI{11.720}{\angstrom}$ along $x$ for both the complexions. In Fig.~\ref{fig:atoms_shuffle}, the initial positions of the atoms in the top and bottom grain before GB migration are plotted in black and their final positions after migration are marked in red and blue for domino and pearl, respectively. At first glance, we can see that the dichromatic pattern appears in the traversed region. This is as expected, due to the rearrangement of atoms in this regions from the crystallography of the top crystal (black, before) to the one of the bottom crystal (red/blue, after) \cite{ZHANG2006623,Race2015}. On closer inspection, it can be seen that there are no true coincidence sites, which is a result of the microscopic degrees of freedom of the GB: The dichromatic pattern is always plotted such that coincidence sites exist, but in reality the top crystal can always be translated arbitrarily against the bottom crystal \cite{Frolov2012a, Frolov2021, Brink_2023}. Furthermore, a more complex pattern arises above/below the traversed region. It appears that atoms do not directly jump from their initial to their final position during one GB migration step. 

Hence, we also analyzed the atomic displacements of perfect $\Sigma19$ GB complexions during a single step of GB migration. A simulation cell of size $11.142 \times 192.291
\times \SI{6.261}{\angstrom^3}$ ($1 \times 10
\times 1$ units cells, \SI{1137}{atoms}) was used for this. We only considered atoms that were not identified as fcc atoms either before or after the GB migration step, utilizing the polyhedral template matching structure identification method \cite{Larsen_2016} as implemented in OVITO \cite{Stukowski2009}. The results are shown in Fig.~\ref{fig:dispmag}. The jump vectors are symmetric around $\lvert \mathbf{b} \rvert /2 = \SI{0.586}{\angstrom} / 2 = \SI{0.293}{\angstrom}$ in $x$ direction. The average of all jumps has to be $\lvert \mathbf{b} \rvert /2$ because the displacement for atoms with $y$ coordinates below the GB has to be zero, while the displacement above the GB has to be $\lvert \mathbf{b} \rvert$. The additional symmetry of the jump vectors is due to the symmetry of the GB. The atomic displacements during a single GB migration step do not correspond to DSC vectors. In our simulations, atoms thus transition from one crystallite to the GB region and only then to the second crystallite during GB migration. The non-DSC nature of the jump vectors is due to the internal degrees of freedom for the atomic positions of the domino and pearl complexions.

We furthermore probed the effort required to effect those jumps by calculating the $L^2$-norm of a combined vector of the displacement components for the GB atoms $i$ as $\sqrt{\sum_{i = 1}^{n}{(x_i^2 + y_i^2 + z_i^2)}}$. The atomic jump lengths are evaluated to be \SI{3.018}{\angstrom} for domino and \SI{2.877}{\angstrom} for pearl. The difference in jump lengths is small and seems to be unlikely to explain the differences in $\tau_c$ for the two complexions. It is therefore necessary to calculate the exact energy cost of introducing the disconnections as in Sec.~\ref{sec:coupling}.

\section{\label{sec:GBcomparison}Shear coupled motion in other GBs of fcc metals in literature}

The energy of disconnections depends on both their core energy and their system-size-dependent elastic interaction energy, and is therefore best described by the parameters $K$ and $\delta_0$ (see Eqs.~\ref{eq:Edipole}--\ref{eq:E_max}). 

Here, $K$ encodes the elasticity of the crystal lattice and $\delta_0$ the properties of the disconnection core. Only a limited number of studies list such values, and for copper GBs we just found Ref.~\cite{Rajabzadeh2013}, in which a $\Sigma13$ $[001]$ $(320)$ symmetric tilt GB is simulated. They obtained a value of $K = \SI{30.468}{meV/\angstrom^3}$ \cite{Rajabzadeh2013}, whereas we found $K = \SI{76.822}{meV/\angstrom^3}$ for $\Sigma$19b and $K = \SI{79.214}{{meV/\angstrom^3}}$ for $\Sigma7$. The difference could be a result of the anisotropy of copper and the different GB planes. The ratio of $K$ values for orientations along the $\langle111\rangle$ and $\langle100\rangle$ directions is 2.517, which is consistent with the ratio of anisotropic Young's moduli for orientations along the $\langle111\rangle$ and $\langle100\rangle$ directions being 2.893 \cite{Freund_Suresh_2004,Armstrong2009}.  The paper reported $E_\text{core} = \SI{5.3}{meV/\angstrom}$ and $\delta_c = \SI{3.615}{\angstrom}$ \cite{Rajabzadeh2013}. The latter was chosen arbitrarily and we therefore combined these values into $\delta_0 = \SI{2.556}{\angstrom}$ (see Eq.~\ref{eq:E_tot_short} and surrounding discussion), which is of the same order of magnitude as our values of $\delta_0 = 3.8$--$\SI{4.4}{\angstrom}$ obtained in $\Sigma$19b GBs and quite lower than values $\delta_0 = 5.082$--$\SI{92.652}{\angstrom}$ obtained in $\Sigma7$ GBs (Table~\ref{tab:fitparameters}). The migration barrier $E_\text{mig}$ was reported as $5.2\pm\SI{0.4}{meV/\angstrom}$ \cite{Rajabzadeh2013}, which is in the same range as our values of $E_\text{mig} = 1.1$--\SI{7.6}{meV/\angstrom} of $\Sigma19$ GB and $E_\text{mig} = 0.532$--\SI{3.430}{meV/\angstrom} of $\Sigma7$ GB.

Previously, the critical shear stress $\tau_c$ for GB migration was calculated for various GBs in fcc metals. Values lie in the range of 1--\SI{4}{GPa} \cite{Rajabzadeh2013, Frolov2014,Larranaga2020,Combe2021}. At \SI{0}{K}, shear stress in $\Sigma13$ $[001]$ $(320)$ and $\Sigma17$ $[001]$ $(410)$ symmetric tilt Cu GBs is observed to be \SI{1.4}{GPa} and \SI{2.1}{GPa}, respectively \cite{Rajabzadeh2013, Combe2021}. Likewise, shear stress in the $\Sigma41$ $[001]$ $(540)$ Al GB is noted to be \SI{2.85}{GPa} \cite{Larranaga2020}. This is in the same range as our \SI{0}{K} values, which are \SI{1.117}{GPa} and \SI{0.849}{GPa} for pearl and domino respectively. Shear stress as a function of temperature is reported for complexions in Cu $\Sigma5$ $[001]$ $(210)$ GB \cite{Frolov2014}. At \SI{500}{K}, shear stress is observed to be $\approx \SI{0.95}{GPa}$ and  $\approx \SI{0.58}{GPa}$ for split kite and filled kite complexions, respectively. At the same temperature, the critical shear stress in this study is $\approx \SI{0.5}{GPa}$ and $\approx \SI{0.4}{GPa}$, respectively. The shear stress difference between the complexions are irrespective of the activation of different disconnection modes. In Cu, the complexions split kites and filled kites of $\Sigma5$ is similar to domino and pearl of $\Sigma7$ in having different disconnection modes contrary to domino and pearl of $\Sigma19$ GB having the same disconnection mode.

\end{document}